\documentclass[fleqn,usenatbib]{mnras}

\usepackage{newtxtext,newtxmath}
\usepackage[T1]{fontenc}

\DeclareRobustCommand{\VAN}[3]{#2}
\let\VANthebibliography\thebibliography
\def\thebibliography{\DeclareRobustCommand{\VAN}[3]{##3}\VANthebibliography}

\usepackage{graphicx}	% Including figure files
\usepackage{amsmath}	% Advanced maths commands
\usepackage{bm}
\usepackage{caption}    % Subfigures
\usepackage{subcaption} % Subfigures
\newcommand{\loss}{\mathcal{L}}

\newcommand{\diff}{\mathrm{d}}
\newcommand{\der}[2]{\frac{\diff#1}{\diff#2}}

\newcommand{\parder}[2]{\frac{\partial#1}{\partial#2}}

\newcommand{\grad}[1]{\bm{\nabla} #1}
\newcommand{\diver}[1]{\bm{\nabla} \cdot #1}
\newcommand{\curl}[1]{\bm{\nabla} \times #1}

\title[GR magnetospheres with PINNs]{General-relativistic magnetar magnetospheres in 3D with physics-informed neural networks}

\author[P. Stefanou et al.]{
Petros Stefanou,$^{1}$\thanks{E-mail: petros.stefanou@ua.es}
Arthur G. Suvorov$^{1,2}$,
Jos{\'e} A. Pons$^{1}$
\\
$^{1}$Departament de Física Aplicada, Universitat d'Alacant, Ap. Correus 99, E-03080 Alacant, Spain\\
$^{2}$Theoretical Astrophysics, IAAT, University of T{\"u}bingen, T{\"u}bingen, D-72076, Germany
}

\date{Accepted XXX. Received YYY; in original form ZZZ}

\pubyear{2025}

\begin{document}
\label{firstpage}
\pagerange{\pageref{firstpage}--\pageref{lastpage}}
\maketitle

% Abstract of the paper
\begin{abstract}
Magnetar phenomena are likely intertwined with the location and structure of magnetospheric currents. 
General-relativistic effects are important in shaping the force-free equilibria describing static configurations, though most studies have quantified their impact only in cases of axial symmetry.
Using a novel methodology based on physics-informed neural networks, fully three-dimensional configurations of varying stellar compactness are constructed. Realistic profiles for surface currents, qualitatively capturing the geometry of observed hotspots, are applied as boundary conditions to deduce the amount of free energy available to fuel outburst activity. It is found that the lowest-energy solution branches permit only a $\approx 30\%$ excess relative to current-starved solutions in axisymmetric cases with global twists, regardless of compactness, reducing to $\approx 5\%$ in 3D models with localised spots. Accounting for redshift reductions to their inferred dipole moments from timing data, explaining magnetar burst energetics therefore becomes more difficult unless the field hosts non-negligible multipoles. Discussions on other aspects of magnetar phenomena are also provided.
\end{abstract}


%%%%%%%%%%%%%%%%%%%%%%%%%%%%%%%%%%%%%%%%%%%%%%%%%%

%%%%%%%%%%%%%%%%% BODY OF PAPER %%%%%%%%%%%%%%%%%%

\section{Introduction}

Magnetars are a class of young, highly magnetised, and slowly rotating neutron stars \citep{Thompson1995, Thompson1996}. 
They are observed mainly in X-rays, either through unpredictable, explosive transients (short bursts, outbursts, giant flares) or through persistent, pulsed thermal emissions \citep{Turolla2015, Kaspi2017}. 
Both types of activity are thought to be linked to twisted, magnetospheric substructure \citep{thomp02, Beloborodov2007}. These currents are thought to be sourced by the displacement of field-line footpoints instigated by mechanical stresses applied to the crust. 
Stress accumulation is typically considered a gradual process driven by the interior field evolution, continuing until a critical threshold is breached. 
At this point, stresses are rapidly released (typically on much faster timescales) via abrupt starquakes \cite[e.g.][]{Perna2011,Pons11,franco00}, plastic motions \cite[e.g.][]{Lander2019}, or magnetospheric instabilities \citep{Carrasco2019}.
Observational \citep{Tiengo2013, Younes2022} and theoretical \citep{Beloborodov2009} evidence suggests that twists develop in localised regions, forming magnetospheric loops akin those in the solar corona.

Axisymmetric, force-free (FF) `Grad-Shafranov' equilibria have been studied extensively in the neutron-star context \cite[e.g.][]{Pili2015, Akguen2017}. Recently, a more general treatment employing a full three-dimensional setup was presented in \cite{Stefanou2023} \cite[see also][]{Carrasco2019, Mahlmann2023}. These works establish the basic features in the modelling of magnetar magnetospheres with an important consensus being that, depending on the boundary conditions imposed at the stellar surface and on the toroidal field in particular, it is possible that there is a unique solution, multiple ones, or none at all \cite[see][]{aly84}. In cases where multiple solutions exist, higher-energy equilibria host disconnected magnetic domains (`islands') and it is expected that only the lowest-energy branch is stable \citep{yu12,Mahlmann2019}. In other words, some form of rearrangement must take place if a magnetospheric configuration becomes `overtwisted' such that a stable state is restored after plasmoid ejection. The energy released in these events is channelled to particles and then radiated away, potentially feeding a magnetar outburst \citep{Lyutikov2003,Lyutikov2006, Parfrey2013, Sharma2023a}. In order to explain the observational phenomenology, it is thus crucial to understand what factors could adjust the available energy budget in a twisted magnetosphere.

For compact objects, general-relativistic (GR) effects reshape the solution landscape
and a self-consistent treatment should take them into account. Focussing on the GR effects on the balance of magnetic stresses, \cite{Kojima2017} reports that the energy that can be stored -- relative to the untwisted, potential state -- in a GR magnetosphere can be several times larger than the corresponding flat-space maximum for ultracompact stars \cite[see also][]{yu11,kojima2018}. However, their most energetic models also correspond to (likely) unstable branches containing disconnected magnetic domains, as described above. Nevertheless, even in the more lenient case of lowest-energy configurations, the stored energy surpasses that of flat-space equilibria \emph{for a fixed dipole moment} because the surface field is augmented by GR factors \cite[e.g.][]{pet74}. On the other hand, the spin-down luminosity scales with stellar compactness for the same reason \citep{Rezzolla2004}. From an observational perspective, this means that the surface field, as inferred from timing data, is lower than the Newtonian case and thus even if more energy can be stored in a relative sense, the absolute value available for outburst activity may be \emph{lower} in reality.

In this paper, we reexamine this interplay by constructing self-consistent, three-dimensional magnetospheric configurations in GR using motivated prescriptions for the twist profile. By comparing observed outburst energies to the available free energy using the rescaled spindown luminosity we show, in fact, that \emph{maximum} dipolar excesses alone are borderline insufficient to power some magnetar outbursts for typical masses and radii \cite[notably the 2009 burst from SGR 0418+5729 and the 2004 giant flare from SGR 1806-20;][]{palm05,cz18}. This provides further, indirect evidence for multipolar components in magnetar magnetic fields.

The equations that govern the magnetospheric field are solved using a physics-informed neural network \cite[PINN;][]{Lagaris1998, Raissi2019,Urban2025}. These machine-learning based solvers build approximate solutions to partial differential equations (PDEs) by minimizing the residuals in a large but scattered set of points throughout the computational domain. They have been successfully applied to a vast range of physical applications and recent advances have positioned PINNs on (at least) equal footing with traditional numerical methods in terms of accuracy and efficiency \cite[see][for a comprehensive review]{Karniadakis2021}. For applications such as the one treated in this work, their main advantages compared to traditional numerical methods are their scalability to higher dimensions with a relatively low toll on computational resources, their meshless nature, and their flexibility in imposing boundary conditions and other physical constraints. A PINN solver was successfully applied by \cite{Urban2023} and \cite{Stefanou2023a} to solve for axisymmetric magnetar and pulsar magnetospheres, respectively---extended here to the fully 3D problem in GR. To our knowledge this is the first time that such solutions have been presented and thus, aside from the astrophysical applications detailed above, this work further highlights the practicality of PINN solvers.

This work is organized as follows. In Section~\ref{sec:equations} we present the basic equations that describe the problem and other useful definitions. In Section~\ref{sec:pinns} we give a brief overview of PINNs as a method for solving these equations, highlighting some important aspects and novelties that we employ. We present our results in Section~\ref{sec:results} and discuss their implications in interpreting the magnetar phenomenology in Section~\ref{sec:discussion}. We conclude with some final remarks, limitations and future plans in Section~\ref{sec:conclusions}.

\section{Force-free magnetic fields}\label{sec:equations}

In what follows, we use the usual Schwarzschild spherical coordinates $(r,\theta,\phi)$, except in appendix \ref{app:numerical_details} where we discuss numerical techniques. We denote the stellar radius by $R$, and we normalise magnetic fields in units of the equatorial field strength, $B_*$. 
Furthermore, we set $G = c = 1$ throughout most of this paper so that, for stellar mass $M$, the ratio $M/R$ defines the stellar compactness.

\subsection{Problem setup}\label{sec:ff field}

Deformations due to rotation and hydromagnetic forces can be safely neglected in the metric surrounding a slow\footnote{Even for a period of $P \lesssim 2$~ms, the $tt$-component of the metric changes by at most $\sim 10 \%$ for compact stars with realistic equations of state, as verified by the RNS code \citep{sf95}.} star with a sub-virial magnetic field ($B \ll 10^{18}$~G). The spacetime line element for cases of interest is therefore well-described by the Schwarzschild solution \citep{Raedler2001, Rezzolla2004, Kojima2017} 
\begin{equation} \label{eq:sch}
    ds^2 = -e^{2\nu(r)} dt^2 + e^{2\lambda(r)} dr^2 + r^2d\theta^2 + r^2 \sin^2{\theta} d\phi^2,
\end{equation}
where $e^{\nu(r)} = e^{-\lambda(r)} = \sqrt{1 - \frac{2M}{r}}$ is the redshift factor. We treat electromagnetic fields as test fields over the background \eqref{eq:sch}, ignoring (again negligible for `weak' fields) corrections to the spacetime geometry due to the magnetosphere itself; see \cite{konno99}. 

In the magnetosphere of a magnetar, 
the magnetic force overwhelmingly dominates, allowing all other forces to be neglected: the magnetic pressure far exceeds the plasma pressure, the $e^{\pm}$ charge carriers carry negligible momentum, inertia is irrelevant, and rotation is too slow to play a significant role.
This is the well-known FF regime, which is mathematically described by
\begin{equation}\label{eq:force-free}
    \bm{\nabla} \times (e^{\nu}\bm{B}) = \alpha \bm{B},
\end{equation}
where 
\begin{equation}
    \bm{\nabla} = e^{-\lambda} \parder{}{r} \hat{\bm{\mathrm{e}}}_r + \frac{1}{r}\parder{}{\theta}\hat{\bm{\mathrm{e}}}_\theta + \frac{1}{r \sin{\theta}}\parder{}{\phi}\hat{\bm{\mathrm{e}}}_\phi 
\end{equation} 
is a weighted gradient operator \cite[e.g.][]{Raedler2001} and we work throughout with the magnetic 3-vector, $\bm{B}$, defined by an orthonormal tetrad (i.e., the field as measured by a locally-inertial observer; when writing $B_{r}$, for instance, this should not be confused with the radial component of the magnetic 4-vector).

The above equation states that the current flowing in the magnetosphere must be parallel to the magnetic field, with the FF parameter $\alpha = \alpha(\bm{x})$ dictating their relative strengths.
It is straightforward to show that equation \eqref{eq:force-free}, along with the divergence-free condition of the magnetic field, imply
\begin{equation}\label{eq:B_dot_grad_alpha}
    \bm{B} \cdot \bm{\nabla} \alpha = 0,
\end{equation}
which means that $\alpha$ is constant along magnetic field lines. 
This function $\alpha$ can be thought of as a measure of twist or helicity, often associated with stored magnetic energy that can drive dynamic processes like magnetic reconnection, flares, or coronal mass ejections. Since $\alpha$ is constant along magnetic field lines, the total twist of a field line can be approximated as $\alpha~L$, where $L$ is the length of the line. In that sense, $\alpha$ can be understood as a `twist density' along a given field line.

To fully define the system, boundary conditions must be specified at the stellar surface. The solution is uniquely determined by setting the radial component of the magnetic field along $r=R$ and assigning a value to $\alpha$
at one end of each footprint, for instance, in regions where
$B_r$ is positive.
 \begin{figure}
     \centering
    \includegraphics[width=0.49\textwidth]{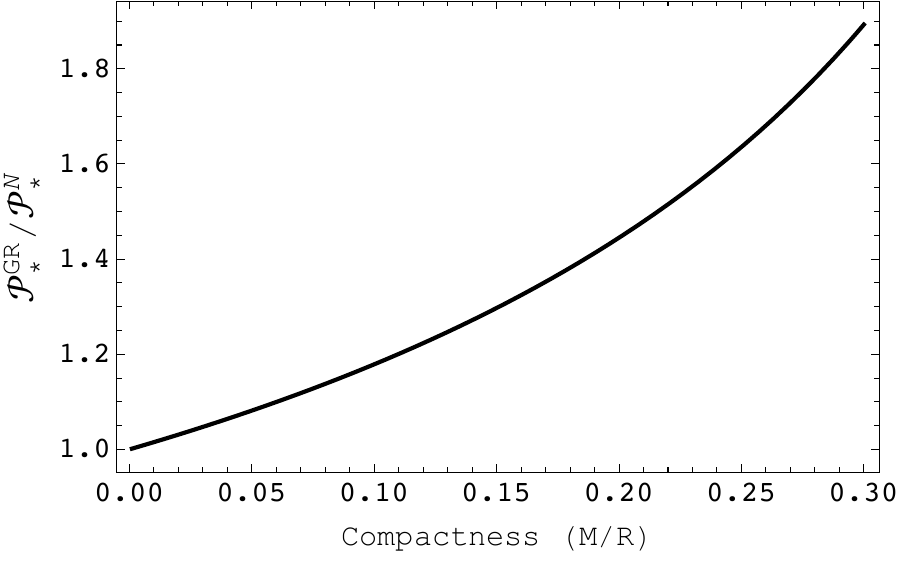}
     \caption{Ratio of the surface, dipolar poloidal fluxes \eqref{eq:P_surface} between the GR and Newtonian ($M=0$) cases as a function of compactness for a fixed dipole moment $\mu_B$.}
     \label{fig:pratios}
 \end{figure}
In principle, both of these boundary conditions encode information about internal processes in the stellar crust, namely its multipolar structure and whether there are any surface currents \cite[for a discussion in an evolutionary context, see][]{Urban2023}.
For the radial magnetic field, it is usually convenient to define a scalar function $\cal P$ such that the poloidal magnetic field is given by $\bm{B}_{\mathrm{pol}} = \grad{\mathcal{P}}\times\grad{\phi}$. 
The scalar function\footnote{Some authors, including \cite{Pons2009}, introduce a different scalar function $\Phi$ such that $\mathcal{P} = r\sin{\theta}\parder{\Phi}{\theta}$ and $\bm{B}_{\mathrm{pol}} = \curl{(\bm{r} \times \grad{\Phi})}$. Furthermore, the reader should also be wary that \cite{Kojima2017} and others define the power law in expression \eqref{eq:alpha_surface_axisymmetric} using a function $S$, related to our $\alpha$ through $\alpha = \der{S}{P}$, reserving the symbol $\alpha$ instead for the timelike component of the metric.} is then expanded in a series of spherical harmonics, with its radial dependence given in terms of Barnes' extended hypergeometric functions with coefficients defining the magnetic multipole moments \cite[e.g.][]{Kojima2017, Pons2009}.

For simplicity, we restrict our attention to models with a surface dipole, given by
\begin{equation}\label{eq:P_surface}
    \mathcal{P}_* = -\frac{3\mu_B R^2}{8M^3} \left[\ln \left(1-\frac{2 M}{R}\right) +\frac{2 M}{R} + \frac{2 M^2}{R^2} \right] \sin^2{\theta},
\end{equation}
where $\mu_B$ is the magnetic dipole moment (as measured by a distant observer) and the subscript~$*$ denotes a restriction to the stellar surface. 
The scaling of expression \eqref{eq:P_surface}, which entails an effective amplification to the magnitude of the magnetic field in the GR case, is depicted in Figure~\ref{fig:pratios}. At fixed $\mu_B$, an increase of $\gtrsim 65\%$ can be seen for very compact stars with $M/R \gtrsim 0.25$, which impacts not only the magnetospheric structure (Sec.~\ref{sec:results}) but also the spindown luminosity (Sec.~\ref{sec:discussion}).
In the absence of magnetospheric currents (i.e. when $\alpha = 0$), equation \eqref{eq:force-free} with boundary condition \eqref{eq:P_surface} can be solved analytically to give the potential solution \cite[e.g.][]{pet74} 
\begin{align} \label{eq:br0}
B_{r_0} &= -\frac{3 \mu_B  }{4 M^3} \left[\ln \left(1-\frac{2 M}{r}\right) +\frac{2 M}{r} + \frac{2 M^2}{r^2}  \right] \cos{\theta}, \\
B_{\theta_0} &=\frac{3 \mu_B}{4 M^3} 
\frac{\left[ \left(1-\frac{2 M}{r}\right) \ln \left(1-\frac{2 M}{r}\right) 
+ \frac{2 M}{r} \left(1-\frac{M}{r}\right) \right]}
{\sqrt{1-\frac{2 M}{r}}} \sin{\theta}, \label{eq:bt0}
\end{align}
and $B_{\phi_0} = 0$.
We refer to this solution as the \textit{untwisted} or \textit{current-free} solution. 
It corresponds to the minimal energy configuration for a given $M$ and $\mu_B$, and serves as a useful reference point to compare twisted fields in either axisymmetric or fully 3D cases.

The other boundary condition, applied to $\alpha$, accounts for currents that may flow in the magnetosphere.
Field lines whose footprints lie in regions where $\alpha \ne 0$ are threaded with currents, develop a toroidal field component and form twisted magnetospheric regions. 
For axisymmetric models, most works assume a power-law dependence of the form
\begin{equation}\label{eq:alpha_surface_axisymmetric}
    \alpha_* = s \mathcal{P}_*^m,
\end{equation}
where $s, m$ are parameters that control the strength of the twist and the non-linearity of the model respectively. The form \eqref{eq:alpha_surface_axisymmetric} is not necessarily motivated by physical considerations but makes for a more mathematically-tractable problem; we use it here to make contact with previous studies (Sec.~\ref{sec:axisym}).

For 3D models (Sec.~\ref{sec:3dresults}), we adopt instead a physically-motivated prescription using the Gaussian profile from \cite{Stefanou2023}, viz. 
\begin{equation}\label{eq:alpha_surface}
    \alpha_* (\theta, \phi) = \alpha_0 \exp\left[\frac{-(\theta - \theta_1)^2 - (\phi - \phi_1)^2}{2 \sigma ^2}\right],
\end{equation}
where $\alpha_0, (\theta_1, \phi_1)$ and $\sigma$ respectively control the magnitude, position, and size of a connected pair of `hotspots'. Although we have no temperature in the model so to speak, it is expected that the dissipation of currents in sub-surface, resistive layers will lead to the formation of localised hotspots which anchor the magnetospheric twist \cite[e.g.][]{Beloborodov2009}. Note that the goal of this work is not to model the formation of these spots; rather, we assume some crustal process has taken place such that they emerge and then investigate how the magnetospheric field, which evolves rapidly relative to the crust, may respond.

\section{Method: PINNs}\label{sec:pinns}

We solve equations \eqref{eq:force-free} and \eqref{eq:B_dot_grad_alpha}, subject to the boundary conditions \eqref{eq:P_surface} and \eqref{eq:alpha_surface_axisymmetric} or \eqref{eq:alpha_surface}, with a PINN solver. 
Here, we give a basic overview of the method.
Our implementation is largely based on that employed in \cite{Urban2023} and \cite{Stefanou2023a}, to which we refer the reader for a more general and in-depth presentation of PINNs as solvers for NS magnetospheres. 

For a set of coordinates $\bm{x}$ in some domain $\mathcal{D}$, a general PDE can be written as 
\begin{equation}\label{eq:general_PDE}
    \Delta u - S(\bm{x}, u) = 0,
\end{equation}
where $\Delta$ is some arbitrary differential operator acting on function $u$ and $S$ is a source term.
The solution $u = u(\bm{x}; \bm{\Theta})$ is approximated by a neural network with adjustable parameters $\bm{\Theta}$ (the weights and biases of the neural network).
A loss function, $\loss$, can be constructed from the residuals of the PDE in some number ($N$, say) of points $\bm{x}_{i} \in \mathcal{D}$.
It is a measure of how well $u$ satisfies equation \eqref{eq:general_PDE},
and is typically defined simply as a type of $L^{2}$ residual,
\begin{equation} \label{eq:loss}
    \loss(\bm{\Theta}) = \frac{1}{N}\sum_{i=1}^N | \Delta u_i(\bm{\Theta})  - S_i(\bm{\Theta})|^2,
\end{equation}
where $u_i(\bm{\Theta}) = u(\bm{x}_i;\bm{\Theta})$ and $S_i(\bm{\Theta}) = S(\bm{x}_i, u_i(\bm{\Theta}))$. 
For physical problems instead described by a \emph{system} of PDEs, the total loss is defined as the sum of the individual losses that correspond to each equation: $\loss = \sum \loss_j$.
The so-called `training' of the network consists of iteratively adjusting its parameters to minimize the loss function. 
In practice, the loss function $\loss$ never reaches exactly zero, and training halts either when the loss falls below a user-specified threshold, indicating convergence, or after a fixed number of iterations if convergence is not satisfactory.

For the case of magnetar magnetospheres, we have a system of four coupled PDEs, given by equations \eqref{eq:force-free} and \eqref{eq:B_dot_grad_alpha} 
To these, we further add the solenoidal condition for the magnetic field $\diver{\bm{B}} = 0$. 
This is necessary, because in regions where $\alpha = 0$ equation \eqref{eq:B_dot_grad_alpha} is trivially satisfied. In Appendix~\ref{app:numerical_details} we give more details about the numerical implementation, including the full explicit expressions of the equations that we solve.

We impose boundary conditions by hard-enforcement \citep[][]{Dong2021, Sukumar2022, Xiao2024, Urban2025}. This means that the desired boundary conditions are implemented on the outputs of the network instead of the loss function and are exactly satisfied by construction (see Appendix~\ref{app:numerical_details}).

%%%%%%%%%%%%%%%%%%%%%%%%%%%%%%%%%%%%%%%%%%%%%%%%%%%%%%%%%%%%%%%%%%%%%%%%%%%%%%%%
\section{Results}\label{sec:results}
%%%%%%%%%%%%%%%%%%%%%%%%%%%%%%%%%%%%%%%%%%%%%%%%%%%%%%%%%%%%%%%%%%%%%%%%%%%%%%%

\begin{figure*}
    \begin{subfigure}[b]{0.33\textwidth}
        \centering
        \includegraphics[width=\textwidth]{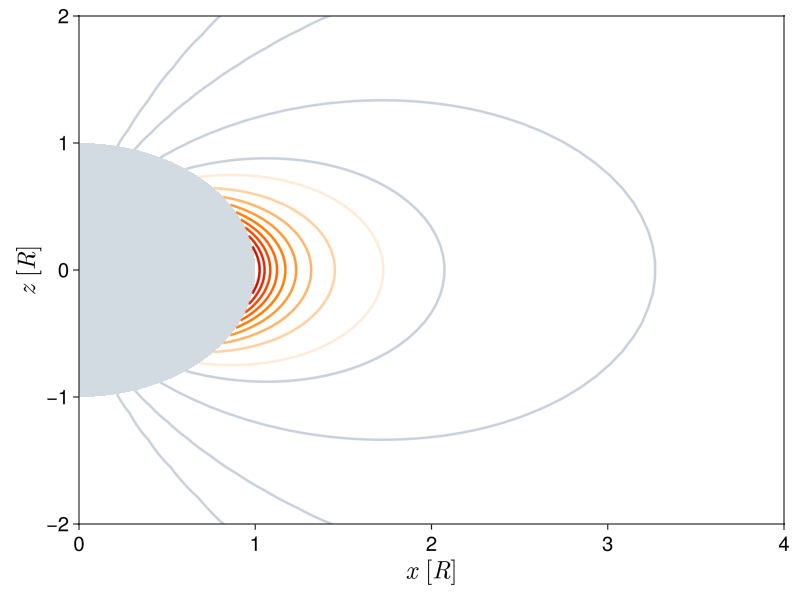}
        \caption{}
        \label{fig:alpha_contour_lines_a}
    \end{subfigure}%
    \begin{subfigure}[b]{0.33\textwidth}
        \centering
        \includegraphics[width=\textwidth]{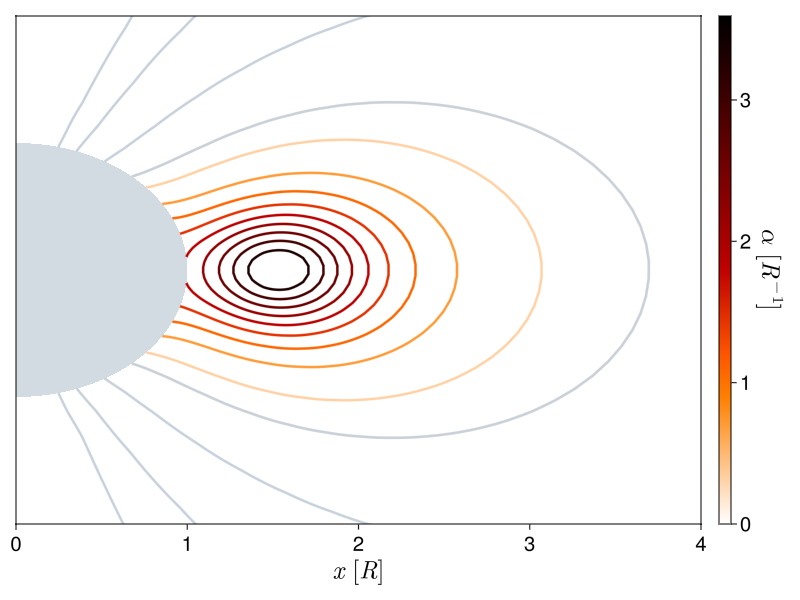}
        \caption{}
        \label{fig:alpha_contour_lines_b}
    \end{subfigure}
    \begin{subfigure}[b]{0.33\textwidth}
        \centering
        \includegraphics[width=\textwidth]{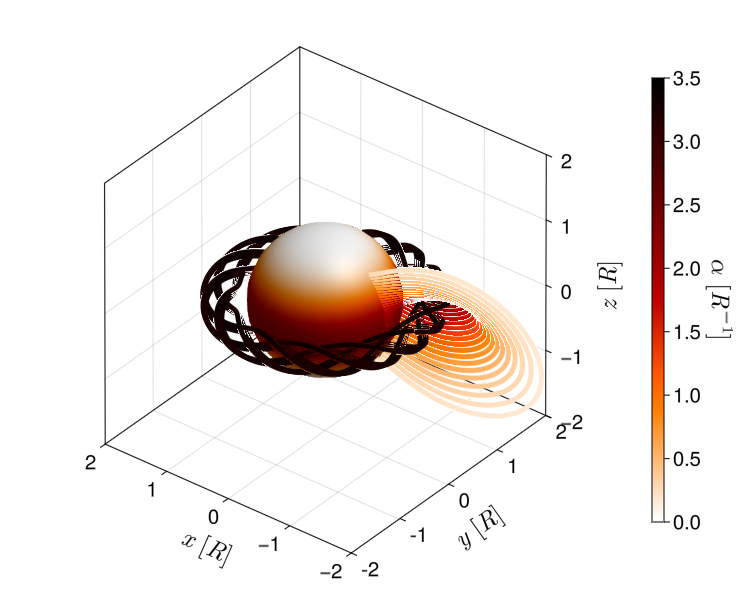}
        \caption{}
        \label{fig:alpha_contour_lines_b}
    \end{subfigure}
    \caption{Isocontours of the twist density $\alpha$ in the $x-z$ plane for axisymmetric solutions for a compact star with $M/R=0.25$, $s = 0.243$ and $m = 3$ (brighter shades indicate greater values of $\alpha$).
    \textbf{(a)}: The low energy solution. \textbf{(b)} The high energy solution, where a detached region (`island') protrudes from the equator. Panel \textbf{(c)} shows a 3D view of the high energy solution, where a disconnected field line with $\alpha \simeq 3.5 R^{-1}$ circles the star. Other twisted lines connected to the surface of the star are also shown for comparison.} 
    \label{fig:alpha_contour_lines}
\end{figure*}

Using the PINN solver described in the previous section, we present some solutions for GR magnetar magnetospheres in this section for either axisymmetric (Sec.~\ref{sec:axisym}) or fully 3D (Sec.~\ref{sec:3dresults}) models. 

While there are several ways one could compare magnetospheric models, a quantity that is particularly useful from an observational perspective is the magnetic energy, $E^{(M)}$; the superscript is used to highlight the implicit dependence on the stellar compactness. To facilitate comparisons between models in the next sections, we introduce the relative excess energy with respect to the untwisted solution %
\begin{equation}\label{eq:excess_energy}
    E_e = \frac{E^{(M)} - E_0^{(M)}}{E_0^{(M)}}.
\end{equation}
Integral expressions for these energies are deferred to Appendix~\ref{sec:magenergy}.

In all cases, numerical accuracy is validated using both the loss function \eqref{eq:loss} and, independently, by comparing equivalent expressions for the energy (Appendix~\ref{sec:magenergy}) establishing the competitiveness of machine-learning techniques with traditional methods.

%%%%%%%%%%%%%%%%%%%%%%%%%%%%%%%%%%%%%%%%%%%%%%%%%%%%%%%%%%%%%%%%%%%%%%%%%%%%%%%%%%%%%%%%%%%%%%%%
\subsection{Axisymmetric models} \label{sec:axisym}
%%%%%%%%%%%%%%%%%%%%%%%%%%%%%%%%%%%%%%%%%%%%%%%%%%%%%%%%%%%%%%%%%%%%%%%%%%%%%%%%%%%%%%%%%%%%%%%%

\begin{figure}
    \centering
    \includegraphics[width=\columnwidth]{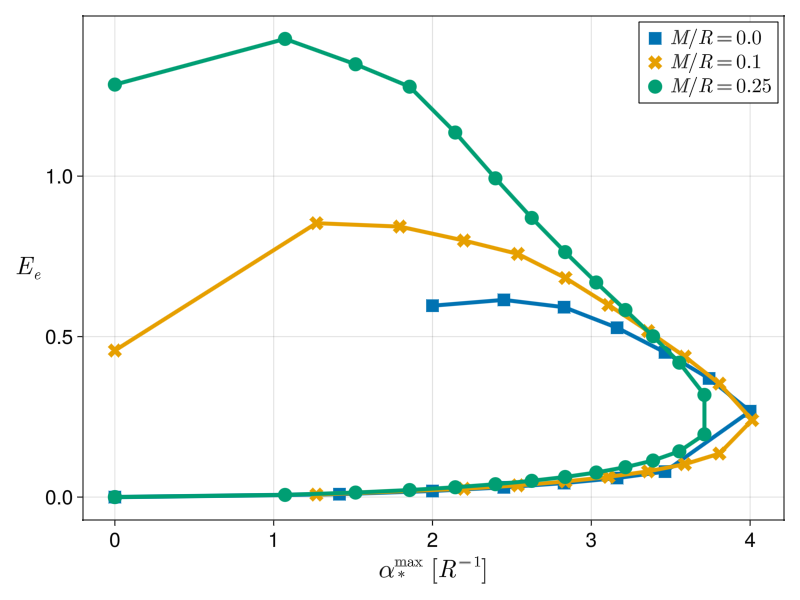}
    \caption{Excess energy $E_e$ as a function of the maximum value of the twist density at the surface $\alpha_*^{\textrm{max}}$ for the axisymmetric models (equation \eqref{eq:alpha_surface_axisymmetric}) for different values of compactness. Two branches of solutions are formed, with different energies for the same model parameters. Beyond a value of $\alpha_*^{\textrm{max}} \approx 4 R^{-1}$, no force-free solutions (stable or unstable) exist.}
    \label{fig:energy_vs_alphamax_axisymmetric}
\end{figure}

We focus first on axisymmetric models. We produce a sequence of magnetospheric solutions for fixed $m=3$ and various values of the parameter $s$ in equation \eqref{eq:alpha_surface_axisymmetric}. Figure \ref{fig:alpha_contour_lines_a} shows an example case with $s = 0.243$ for a compact star with $M/R = 0.25$. Field lines of fixed $\alpha$ are plotted in the $x-z$ plane. We see that the magnetospheric configuration is that of an `inflated dipole', as is characteristic of such `power-law' models from the literature \cite[e.g.,][]{Flyer2004, Pili2015, Akguen2018}. In a more direct sense, the field structure can be compared with the middle panel in figure 3 from \cite{Kojima2017} which also has $M/R = 0.25$ and exhibits the same qualitative features: field lines are pinched near the equator and inflated at larger radii by toroidal pressures, encapsulated by the twist density (reaching a maximum of $\approx 4R^{-1}$ near the surface).

This pinching effect becomes more pronounced as the parameter $s$ increases, leading to a regime where the elliptic equation \eqref{eq:force-free} in its axisymmetric limit admits degenerate solutions with a fully detached `island'. These solutions have the same boundary conditions (i.e. $s$ value) and compactness but differ in energy and magnetospheric structure. One rather extreme case is shown in Figure~\ref{fig:alpha_contour_lines_b}, where a magnetically-disconnected domain is visible around $r \sim 2 R$. We find that the relative energy excess in this case is $E^{(0.25)}_{e} = 1.28$, which can be compared with the `stable' case shown in the left panel which has a free energy that is a factor 60 smaller, $E^{(0.25)}_{e} = 0.02$. We use the word \emph{stable} since \cite{Mahlmann2019} showed that such `upper branch' solutions with islands are dynamically unstable and, therefore, unlikely to be reached in astrophysically relevant scenarios. While islands are not special to GR cases, this huge energy disparity highlights the magnetic confinement effect discussed by \cite{Kojima2017}. For any compactness, increasing the value of $s$ tends to shift the island further away from the star until eventually no solution exists at all, and vice versa.

%or any compactness, increasing (decreasing) the value of $s$ tends to shift the island further (closer) away from the star, until eventually no solution exists at all. ...] and vice versa

\begin{figure*}
    \centering
    \begin{subfigure}[b]{0.33\textwidth}
        \centering        \includegraphics[width=\textwidth]{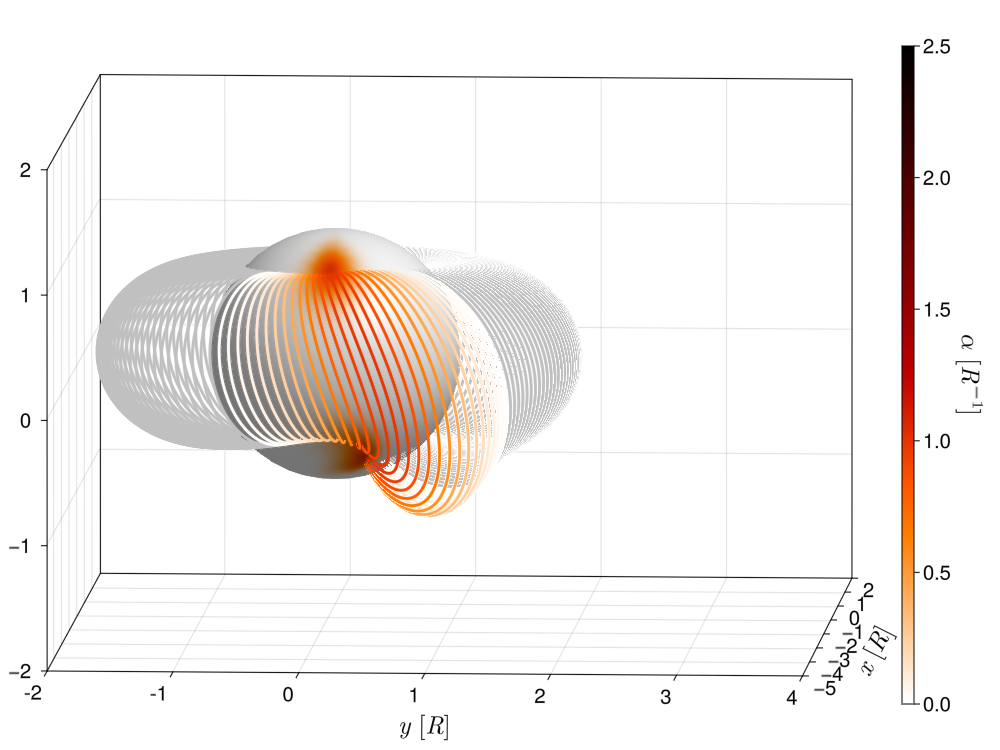}
        \caption{$M/R = 0.0, \ \alpha_*^{\rm max} = 1.0 R^{-1}$}
        \label{}
    \end{subfigure}%
    \begin{subfigure}[b]{0.33\textwidth}
        \centering        \includegraphics[width=\textwidth]{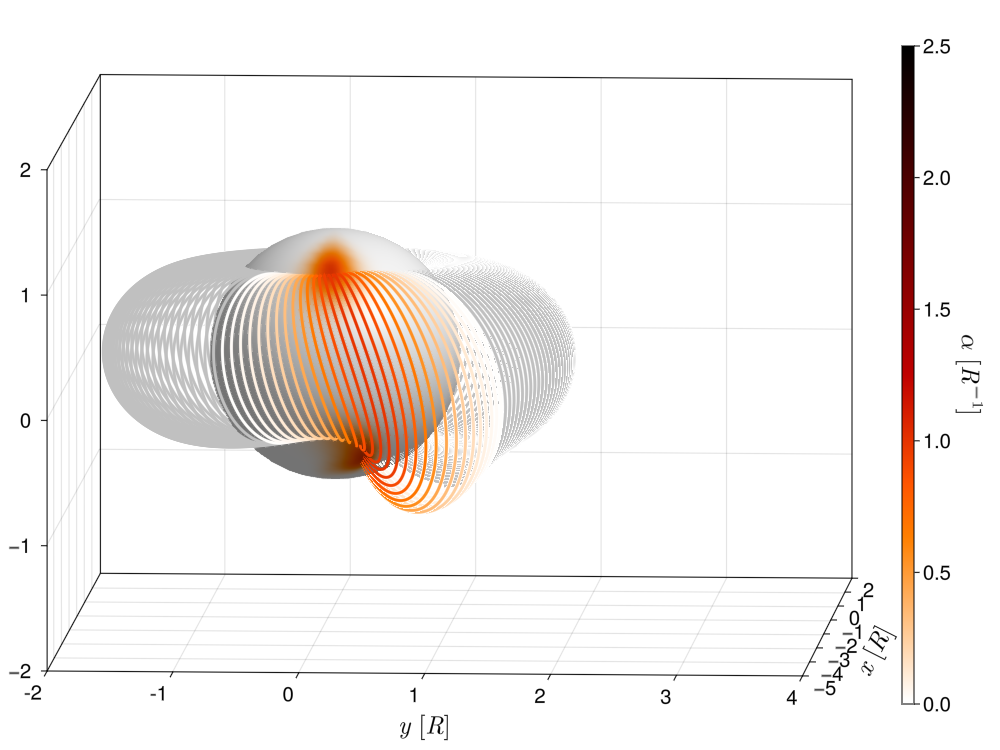}
        \caption{$M/R = 0.1, \ \alpha_*^{\rm max} = 1.0 R^{-1}$}
        \label{}
    \end{subfigure}%
    \begin{subfigure}[b]{0.33\textwidth}
        \centering        \includegraphics[width=\textwidth]{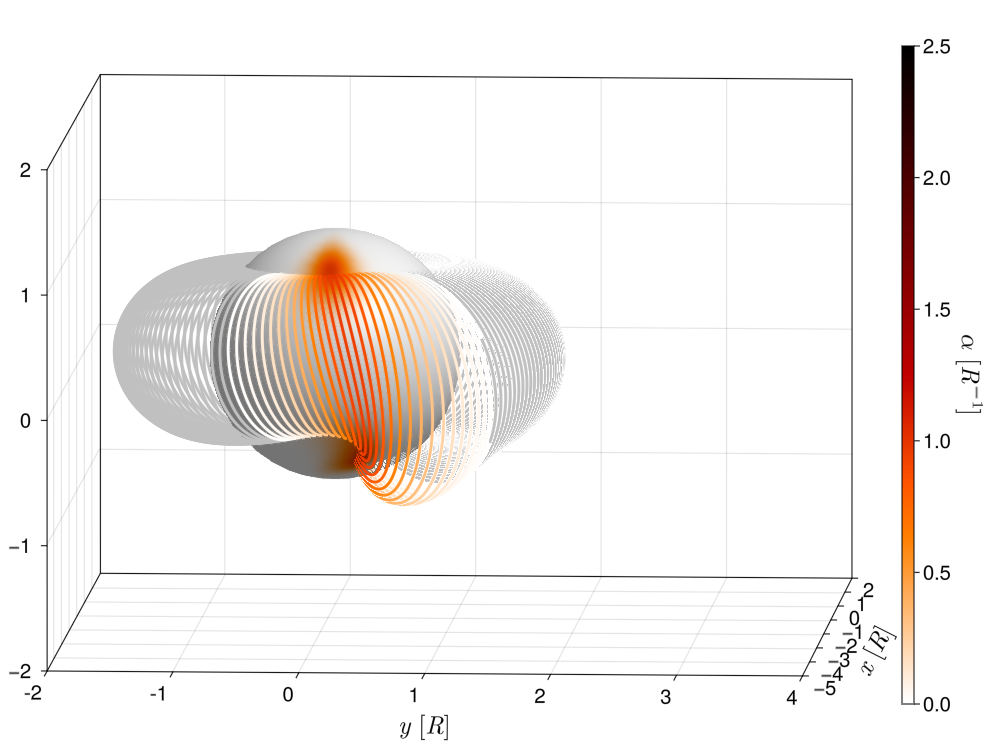}
        \caption{$M/R = 0.25, \ \alpha_*^{\rm max} = 1.0 R^{-1}$}
        \label{}
    \end{subfigure}

    \begin{subfigure}[b]{0.33\textwidth}
        \centering        \includegraphics[width=\textwidth]{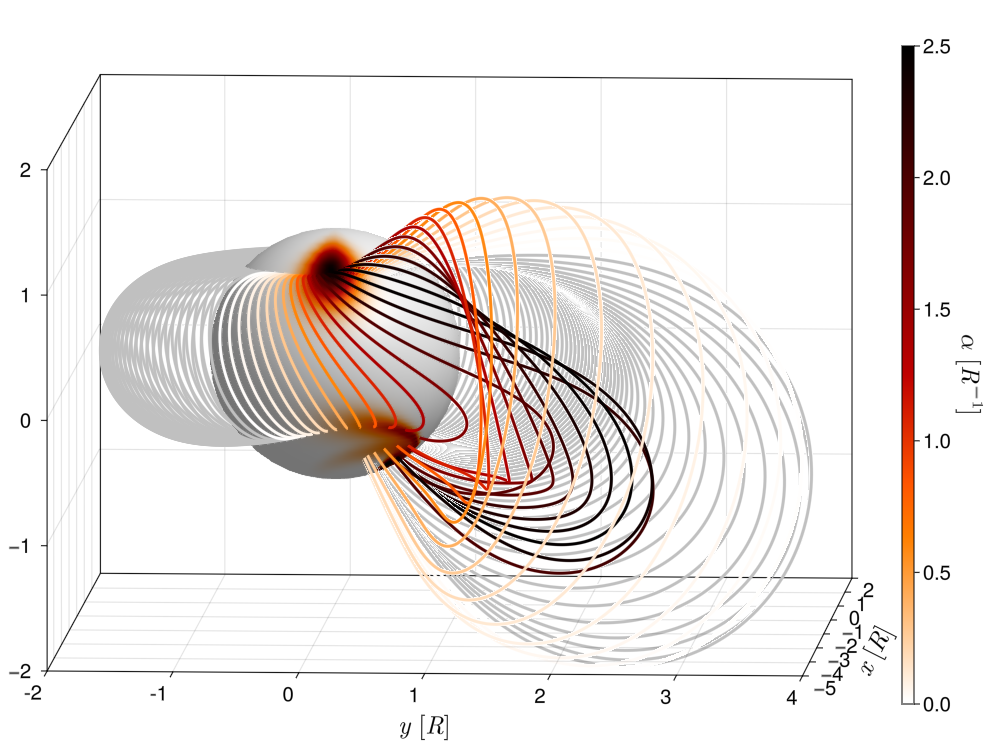}
        \caption{$M/R = 0.0, \ \alpha_*^{\rm max} = 2.5 R^{-1}$}
        \label{}
    \end{subfigure}%
    \begin{subfigure}[b]{0.33\textwidth}
        \centering        \includegraphics[width=\textwidth]{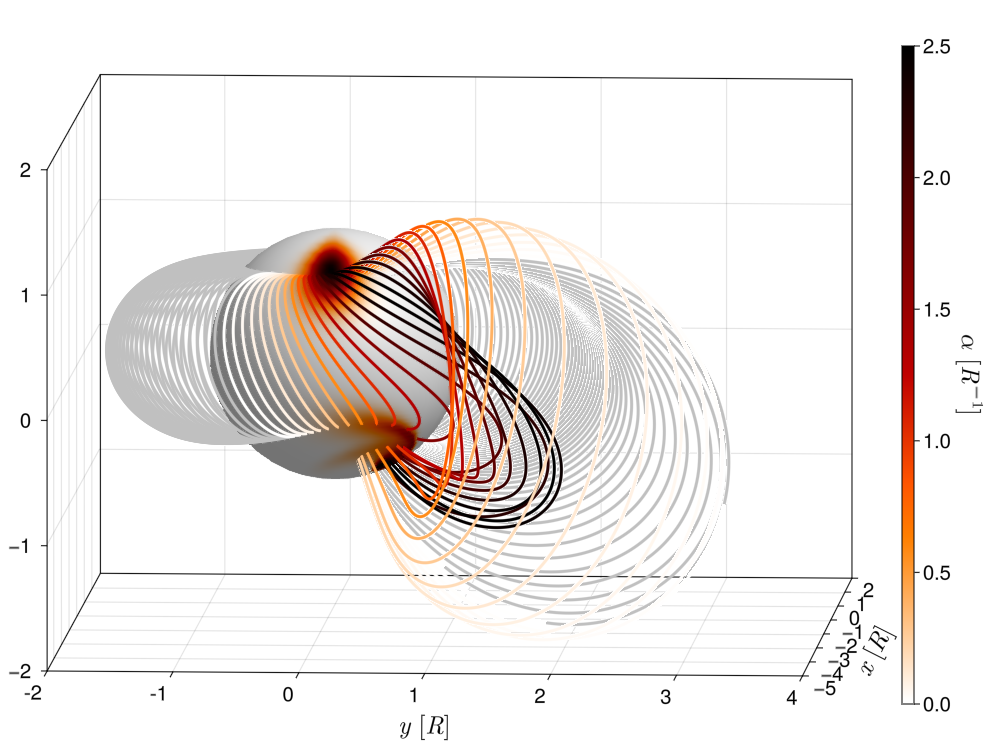}
        \caption{$M/R = 0.1, \ \alpha_*^{\rm max} = 2.5 R^{-1}$}
        \label{}
    \end{subfigure}%
    \begin{subfigure}[b]{0.33\textwidth}
        \centering        \includegraphics[width=\textwidth]{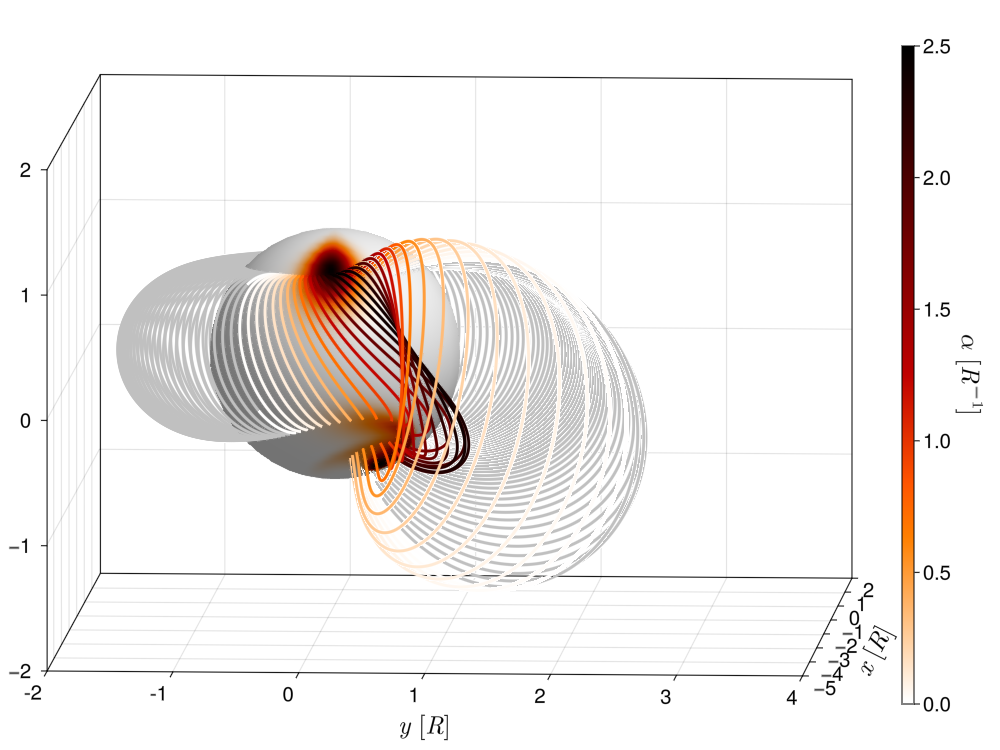}
        \caption{$M/R = 0.25, \ \alpha_*^{\rm max} = 2.5 R^{-1}$}
        \label{}
    \end{subfigure}%
    \caption{Twisted magnetospheres projected into Cartesian-like coordinates for different values of compactness $M/R$ (by column) and current strengths $\alpha_*^{\rm max}$ (by row). Darker shades indicate greater values of $\alpha$, which is constant along any given line.}
    \label{fig:twisted_magnetospheres}
\end{figure*}

Figure~\ref{fig:energy_vs_alphamax_axisymmetric} shows the dependence of the excess energy, $E_e$,  on the maximum value of the twist density at the surface, $\alpha_*^{\mathrm{max}}$ (which depends both on $s$ and $M/R$), for different values of the compactness.
For values of $\alpha_*^{\mathrm{max}} \gtrsim 4 R^{-1}$, we find that no solutions exist for any compactness. 
The sequence has a turning point at this value, bifurcating into two branches with different energies.
In order to facilitate convergence and build sequences of degenerate models, we initialise the solver using the previous converged solution with slightly modified values of the parameters defining the boundary condition.
Physically, the critical point $\alpha_*^{\mathrm{max}} \approx 4 R^{-1}$ corresponds to the most energetic, stable configuration. 
Notably, the excess is largely independent of the compactness and the energy stored in the magnetosphere is $\approx 30\%$ higher than the untwisted case at the turning point. Such a value agrees with \cite{Akguen2018}, who found $E_{e}^{\rm max} \approx 25\%$ in the Newtonian limit. This indicates that, for a fixed dipole moment $\mu_B$, GR magnetospheres can store more energy than their Newtonian counterparts in absolute terms because $E^{(M)}_{0}$ increases monotonically with $M/R$ (though see Sec.~\ref{sec:discussion}). In a relative sense however, the free energy that stable branches can hold is practically independently of compactness.

%%%%%%%%%%%%%%%%%%%%%%%%%%%%%%%%%%%%%%%%%%%%%%%%%%%%%%%%%%%%%%%%%%%%%%%%%%%%%%%%
\subsection{3D models} \label{sec:3dresults}
%%%%%%%%%%%%%%%%%%%%%%%%%%%%%%%%%%%%%%%%%%%%%%%%%%%%%%%%%%%%%%%%%%%%%%%%%%%%%%%%

While useful for estimating the general properties of twisted magnetospheres, axisymmetric models are rather limited with respect to explaining observational phenomenology. For this, a 3D approach with localised twisted regions is more suitable \cite[see][]{Beloborodov2009}.
Such models were studied in \cite{Stefanou2023} in the Newtonian limit. Here, we generalise their results to include relativistic effects using the novel PINN solver described in Sec.~\ref{sec:pinns}. 

We produce sequences of magnetospheric solutions varying the parameter $\alpha_0$ in equation \eqref{eq:alpha_surface} which controls the overall strength of the twist. To avoid scanning a huge parameter space, the rest of the parameters remain fixed to the values $\theta_1 = \frac{\pi}{4}, \phi_1 = \pi, \sigma=0.2$. These numbers roughly describe a spot of radius $\approx 2$ km. 

To understand the impact of the stellar compactness, we study three families of models with $M/R = 0, 0.1, 0.25$. Example configurations for two different values of $\alpha_*^{\rm max}$ (which coincides with $\alpha_0$ in this case, see equation \ref{eq:alpha_surface}) are shown in Figure~\ref{fig:twisted_magnetospheres}, corresponding to either a mild ($\alpha_*^{\rm max} = 1.0 R^{-1}$; top) or extreme ($\alpha_*^{\rm max} = 2.5  R^{-1}$; bottom panels) twist with increasing compactness going from left to right. We see that, much like in the axisymmetric case, the toroidal component of the field exerts a magnetic pressure which `inflates' the magnetosphere in the current-filled region for larger $\alpha_*^{\rm max}$ noting that the scale extends to $r \sim 4 \ R$. The equatorial extent of the twisted flux bundle protruding from the spots decreases monotonically as a function of compactness (by a factor $\sim 1.5$ comparing the $M/R=0.25$ and $M=0$ cases), which further softens the effect of the twist. Because the field is locally stronger near the surface in cases with higher compactness (equation~\ref{eq:P_surface}), the magnetic pressure is larger for a given $\mu_B$ which may be responsible for this `shrinking' effect. In all cases, we see that only the field lines that emerge from the hotspots are twisted, which can be contrasted with the axisymmetric case where twists are always `global' in some sense. Larger values of $\alpha_*^{\rm max}$ lead to more twisted configurations as expected, since the currents are stronger in this case. For example, comparing the final column we see a `weaving' pattern only in in the larger twist case and that the toroidal blowout is a factor $\lesssim 2$ more effective.

\begin{figure}
    \centering
    \includegraphics[width=\linewidth]{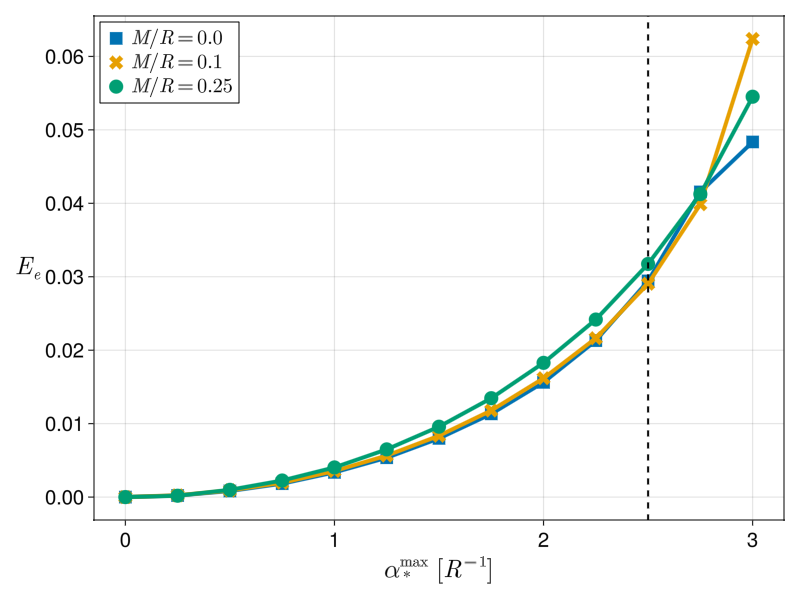}
    \caption{Excess energy $E_e$ as a function of the maximum value of the twist density at the surface $\alpha_*^{\textrm{max}}$ for the 3D models (equation \eqref{eq:alpha_surface}) for different values of compactness. The vertical dashed line marks the value of $\alpha_*^{\textrm{max}} \simeq 2.5 \ R^{-1}$ beyond which the precision degrades, likely associated with the non-existence of solutions. (see Appendix \ref{sec:magenergy}).}
    \label{fig:energy_vs_alphamax_3d}
\end{figure}

In Figure~\ref{fig:energy_vs_alphamax_3d} we show the relative excess energy $E_e$ as a function of $\alpha_*^{\mathrm{max}}$. Interestingly, neither the functional dependence of the energy on $\alpha_*^{\rm max}$ nor the maximum achievable twist seem to be affected by the compactness. In particular, though a one-to-one comparison is made difficult by the different choices of current function, these features were also observed in \emph{stable} axisymmetric cases (Fig.~\ref{fig:energy_vs_alphamax_axisymmetric}). Quantitatively however, because the twist is more localised the excess energy is lower.
Note, however, that models with higher $M /R$ have higher absolute energy for a \emph{fixed} dipole moment because the surface field is effectively amplified (Fig.~\ref{fig:pratios}).

Models with high ($\gtrsim 2.5 R^{-1}$) values of $\alpha_*^{\rm max}$ correspond to extreme configurations that are likely close to or even beyond some physical limit of existence and, thus,  challenge the numerical solver. We refer to
Appendix~\ref{sec:magenergy} for a detailed discussion about this issue.

%%%%%%%%%%%%%%%%%%%%%%%%%%%%%%%%%%%%%%%%%%%%%%%%%%%%%%%%%%%%%%%%%%%%%%%%%%%%%%%%%%%%%%%%%%%%%%%%
\section{Connections to magnetar phenomena}\label{sec:discussion}
%%%%%%%%%%%%%%%%%%%%%%%%%%%%%%%%%%%%%%%%%%%%%%%%%%%%%%%%%%%%%%%%%%%%%%%%%%%%%%%%%%%%%%%%%%%%%%%%

Having constructed sequences of twisted magnetospheres, we consider some implications with respect to observational phenomena.

%%%%%%%%%%%%%%%%%%%%%%%%%%%%%%%%%%%%%%%%%%%%%%%%%%%%%%%%%%%%%%%%%%%%%%%%%%%%%%%%%%%%%%%%%%%%%%%%
\subsection{Spindown luminosity and surface-field inferences} \label{sec:sdlum}
%%%%%%%%%%%%%%%%%%%%%%%%%%%%%%%%%%%%%%%%%%%%%%%%%%%%%%%%%%%%%%%%%%%%%%%%%%%%%%%%%%%%%%%%%%%%%%%%

In the vicinity of a compact star, there is an effective amplification of the magnetic field strength near the pole due to spacetime curvature, as can be read off from expression \eqref{eq:br0}:
\begin{equation} \label{eq:relb}
    \frac{B_{r_0}}{B_{r_0}(M \to 0)} =  -\frac{3}{8} \left(\frac{R}{M}\right)^3 \left[\ln{\left(1 - \frac{2 M}{R} \right)} + \frac{2 M}{R} + \frac{2 M^2}{R^2} \right].
\end{equation}
Furthermore, the angular velocity of the star, as measured by a distant observer, is rescaled by a gravitational redshift equal to 
\begin{equation}
   \Omega = \sqrt{1 - \frac{2 M}{R}} \Omega_*,
\end{equation}
where $\Omega_*$ is the star-frame velocity. These facts together imply that the spindown luminosity, $L$, increases with compactness, as established by \cite{Rezzolla2004}. 
In particular, since $L \propto B_*^2 \Omega_*^4$ there is an effective amplification relative to the Newtonian luminosity by a factor $f_R^2/N_R^4$, where $f_R$ is the right-hand side of expression \eqref{eq:relb} and $N_{R} = e^{\nu(R)} = \Omega/\Omega_*$ \cite[see equation 150 in][]{Rezzolla2004}: 
\begin{equation} \label{eq:sdlum}
 \frac{L^{\rm GR}}{L^{\rm N}} =  \kappa = f_R^2/N_R^4.
\end{equation} 
 This ratio, $\kappa$, scales sharply with compactness. For example, $\kappa = 4.2$ for $M/R = 0.17$ but $\kappa = 10.7$ for $M/R = 0.25$ (see also Sec.~\ref{sec:afterglows}).

The equation obtained by matching $L^{\rm GR}$ to the rotational kinetic-energy loss can be rearranged to produce an expression for the (untwisted) polar field strength. Further correcting for the fact the star ought to be an oblique rotator surrounded by magnetospheric plasma rather than in vacuum \citep{petri16}, we have (restoring units)
\begin{equation} \label{eq:breduc}
\begin{aligned}
B_{p}^2 &= \frac{1}{\kappa \left(a + b \sin^2 \chi \right)} \times \frac{3 c^3 P \dot{P} I_{0}}{2 \pi^2 R^6}\\
&\approx \left(1 - \frac{2 G M }{c^2 R} \right)^{7/2} \frac{3 c^3 P \dot{P} I_{0}}{2 \pi^2 R^6 \left(a + b \sin^2 \chi \right)},
    \end{aligned}
\end{equation}
which one may recognize as the \cite{Spitkovsky2006} formula multiplied by a redshift prefactor for (GR-corrected) moment of inertia $I_{0}$ and magnetic inclination angle $\chi$. 
The second line in equation \eqref{eq:breduc} is simpler to use in practice and valid to within $< 3\%$ for $M/R \leq 0.2$. The factors $a$ and $b$ are fitting coefficients of order unity; in fact they are also generally \emph{larger} than their Newtonian counterparts \cite[see Table 2 in][]{petri16} indicating a further reduction. 
In the following, we take $a = 1$ and $b=1.5$ and use the GR Tolman-VII moment of inertia, $I_{0} \approx 2 M R^2 (1 - 1.1 M/R - 0.6 M^2/R^2)^{-1}/7$ \citep{latprak01}, which is larger than the Newtonian value ($I_{0}^{N} = 2 M R^2/7$).

For a twisted magnetosphere, the Poynting flux may be amplified further as toroidal pressures inflate field lines through the light cylinder \citep{Parfrey2013}. 
For example, \cite{ng25} found that the spindown luminosity may increase by a factor $\lesssim 16$ for extreme twists. Although their results cannot be directly applied to our case owing to different choices for the current function and spin periods of interest, taken at face value this would indicate a further decrease in the inferred polar field strength by a factor $\lesssim 4$. Persistent twist injections may help to explain how some strong-field objects appear to be able to reach long periods \citep{sdp25}.

In the interests of providing a simple, ready-to-use formula using `canonical numbers' ($M=1.4M_{\odot}, R = 12 \text{ km}, \chi = \pi/4$), one finds that the estimate \eqref{eq:breduc} reads
\begin{equation} \label{eq:bpgr}
    B^{\rm GR}_{p} \sim 1.6 \times 10^{19} \sqrt{ P \dot{P}} \text{ G},
\end{equation}
which is notably smaller than the often-quoted Newtonian, vacuum value of
\begin{equation} \label{eq:bpnewt}
    B^{\rm N}_{p} \sim 6.4 \times 10^{19} \sqrt{ P \dot{P}} \text{ G}.
\end{equation}
Note that expression \eqref{eq:bpgr} should apply to the whole neutron-star population and not just magnetars.

\begin{figure}
     \centering
    \includegraphics[width=0.49\textwidth]{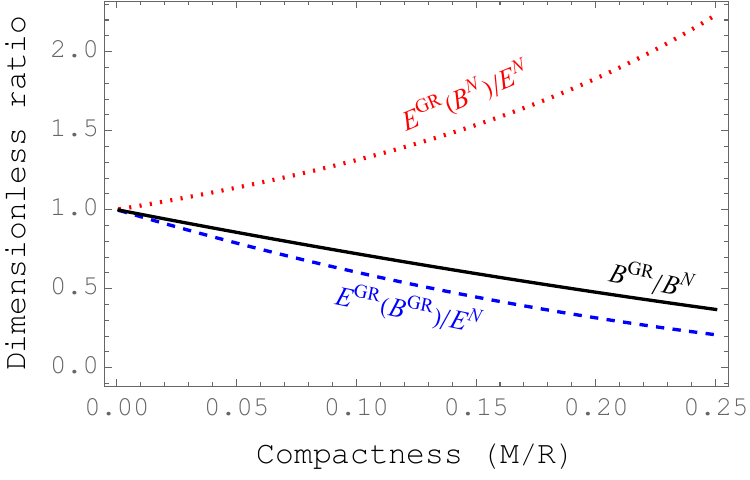}
     \caption{Ratios of dipolar magnetospheric energies normalised by the Newtonian values using the (GR or Newtonian) Tolman-VII equation of state as a function of compactness, with either fixed $\mu_B$ (dotted) or \textbf{$P \dot{P}$} (dashed). The latter applies to cases where the field is inferred from timing data, with the ratio of surface-field inferences depicted by the solid line (see text).}
     \label{fig:ratios}
 \end{figure}

Figure~\ref{fig:ratios} displays some relevant dimensionless ratios, as functions of compactness, assuming the factors $a$ and $b$ in expression \eqref{eq:breduc} are independent of $M/R$ and spin for simplicity \cite[cf.][]{ruiz14,petri16}. 
The top (dotted) line shows the results for the ratio of GR to Newtonian dipolar magnetospheric energy, assuming a fixed dipole moment $\mu_B$. The energy is obtained using equation \eqref{eq:e0}.
% \textbf{ The top (dotted) line shows the numerical results for the ratio of GR to Newtonian magnetospheric energy, assuming a fixed dipole moment. The energy is obtained by numerically integrating the solution over the domain volume.}
For example, for a star with $M/R = 0.2$ the total energy is roughly doubled because the poloidal flux $\mathcal{P}$ is larger than the Newtonian value by a factor $\approx 1.4$ (see Fig.~\ref{fig:pratios}). 
In contrast, the bottom (dashed) line shows the same ratio, 
but for cases where the surface field is normalized under the assumption of a fixed $P \dot{P}$. 
The ratio $B_p^{GR}/B_p^{N}$ is deduced from equation \eqref{eq:breduc} and is depicted by the solid line in the same Figure.

We see that while the energy itself scales with compactness (as per the dotted line), it does so at a rate that is \emph{shallower} than the spindown luminosity \eqref{eq:sdlum}. 
This implies that magnetar magnetospheres house \emph{less} energy in the GR case if the field strength is deduced from timing parameters rather than assumed. For a canonical compactness of $M/R = 0.17$, for example, the total magnetospheric energy is $\approx 38\%$ of the Newtonian value for any given $P \dot{P}$ (dashed line). The key conclusion is that since we find that the available free energy is almost constant as a function of compactness (see Figs.~\ref{fig:energy_vs_alphamax_axisymmetric} and \ref{fig:energy_vs_alphamax_3d}), a decrease by a factor $\sim 3$ in the (timing-inferred) energy available to fuel outbursts is expected for a typical neutron star ($M/R \sim 0.17$).

%%%%%%%%%%%%%%%%%%%%%%%%%%%%%%%%%%%%%%%%%%%%%%%%%%%%%%%%%%%%%%%%%%%%%%%%%%%%%%%%
\subsection{Outburst energetics} \label{sec:outbursts}
%%%%%%%%%%%%%%%%%%%%%%%%%%%%%%%%%%%%%%%%%%%%%%%%%%%%%%%%%%%%%%%%%%%%%%%%%%%%%%%%

 \begin{figure*}
     \centering
    \includegraphics[width=0.97\textwidth]{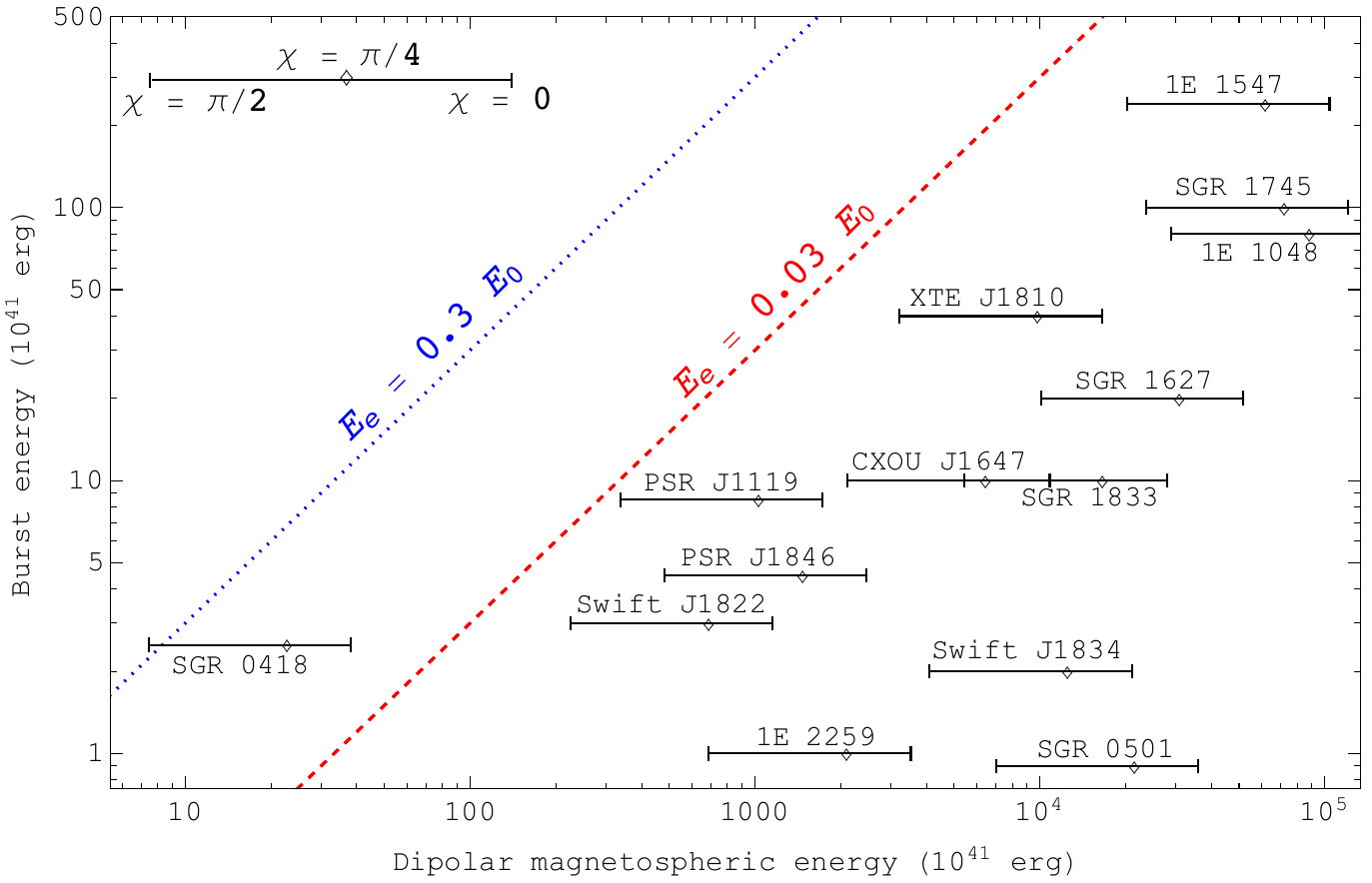}
     \caption{Comparison between \emph{dipolar} magnetospheric energies (see Appendix~\ref{sec:magenergy}) and burst energies from the catalogue of \protect\cite{cz18} for a number of sources (see overlaid legends). In cases of multiple bursts by a given object, only the brightest event is shown. The polar-dipole strengths are determined using the GR-revised relation \eqref{eq:breduc} via the available timing data for masses $M = 1.4  M_{\odot}$, radii $R = 12$~km, Tolman-VII moment of inertia, and inclinations spanning $0 \leq \chi \leq \pi/2$: that corresponding to the `canonical' value \eqref{eq:bpgr} is depicted by small rhomboids (see the top-left legend). The dotted, blue line depicts the maximum free energy available in cases with axisymmetric twists ($\approx 30\%$ of the total; Fig.~\ref{fig:energy_vs_alphamax_axisymmetric}), with the dashed, red line corresponding instead to cases with localised twists ($\approx 3\%$; Fig.~\ref{fig:energy_vs_alphamax_3d}).}
     \label{fig:freevsburst} 
 \end{figure*}

Expression \eqref{eq:breduc} can be used to determine the free energy in the magnetosphere in terms of observables. Recalling Fig.~\ref{fig:ratios}, we have that the total magnetospheric energy is lower by a factor $\sim 3$ relative to the  timing-based Newtonian inference for typical masses and radii ($M/R \approx 0.17$). This prompts us to revisit magnetar burst energies, $E_{\rm burst}$, in a GR context.

Figure~\ref{fig:freevsburst} shows $E_{\rm burst}$ values versus $E^{(0.17)}_{0}$ for several sources using data collated in the catalogue of \cite{cz18}. The reported timing parameters ($P\dot{P}$) are used to estimate the GR-corrected polar-field strength using equation \eqref{eq:breduc}, from which we calculate the total energy (expression~\ref{eq:e0}). Typical values of the energy are found by applying expression \eqref{eq:bpgr}, as shown by the small rhomboids ($\chi = \pi/4)$, though we allow for a range of inclinations ($ 0 \leq \chi \leq \pi/2$) which define horizontal `error bars'. In particular, orthogonal rotators spindown faster for a fixed dipole moment meaning that matching to a given $P$ and $\dot{P}$ implies a lower magnetospheric energy relative to inferences for $\chi = \pi/4$. We overlay lines in the Figure to show the maximum free energies for stable cases with axisymmetric ($E_{e} = 0.3$; Fig.~\ref{fig:energy_vs_alphamax_axisymmetric}) or localised ($E_{e} = 0.03$; Fig.~\ref{fig:energy_vs_alphamax_3d}) twists.

Even accounting for GR reductions to $E^{(M)}_{0}$ and assuming orthogonal rotators, we see that the maximum free-energy for localized twists is sufficient to fuel most outbursts (i.e., $E_{\rm burst} < 0.03 E^{(0.17)}_{0}$). One event, however, which could\footnote{Note the distance of this source ($\sim 2$~kpc) is set mainly by assuming it resides within the Perseus arm. This means the burst energy could be different by a factor few or more in reality if the source is further away or closer, especially when considering beaming uncertainties \cite[see][]{guil15b}.} require $E_{\rm burst} > 0.03 E^{(0.17)}_{0}$ comes from the `low-field' magnetar SGR 0418+5729 \citep{van10}: the release of $E_{\rm burst} \sim 2 \times 10^{41}$~erg in the 2009 event \citep{rea13} points towards multipolarity since $E^{(0.17)}_{0} \sim 3 E_{\rm burst}$ if the inclination angle $\chi \approx \pi/2$ as inferred from the double-peaked light curves observed by the Rossi X-ray Timing Explorer \cite[RXTE;][]{esp10}. Even in a conservative scenario with $\chi = 0$, if $\sim 3\%$ of the total energy was released then $l > 1$ multipoles would have to comprise $> 2$ times more energy than the dipole. For other cases, such as for the 2003 outburst from XTE J1810--197 ($E_{\rm burst} \approx 4 \times 10^{42}$~erg), a dipolar free energy of $< 2\%$ would suffice.

The most energetic flare recorded from a (Galactic) magnetar came from the December 2004 event from SGR 1806--20, with a total (isotropic) energy yield of up to $\sim 5 \times 10^{46}$~erg depending on distance \citep{palm05,tere05}. For an orthogonal rotator, the observed period and period derivative values yield a polar field estimate of $B_{p} \approx 7 \times 10^{14}$~G, a factor $\sim 5$ smaller than the Newtonian, vacuum value ($B_{p} \approx 4 \times 10^{15}$~G) from expression \eqref{eq:bpnewt}. The maximum free energy in the magnetosphere can be estimated as
\begin{equation} \label{eq:sgr}
    E^{1806}_{\rm free} \approx 3.9 \times 10^{46} \left( \frac{B_{p}}{7 \times 10^{14} \text{ G}} \right)^2 \left( \frac{E_{e}}{0.3} \right) \text{ erg} .
\end{equation}
Thus, although smaller distances, beaming, smaller inclination angles, or a less compact star could be invoked to shift the observed energy below that of expression \eqref{eq:sgr}, multipolar components may be anticipated to explain the energetics. At least in the axisymmetric context, \cite{Akguen2018} explored twist profiles different to our expression \eqref{eq:alpha_surface_axisymmetric} and established a free-energy upper limit of $\approx 25\%$ for stable cases, so that we expect $E_{e} = 0.3$ to be a firm bound.

%%%%%%%%%%%%%%%%%%%%%%%%%%%%%%%%%%%%%%%%%%%%%%%%%%%%%%%%%%%%%%%%%%%%%%%%%%%%%%%%
\subsection{Afterglow light curves} \label{sec:afterglows}
%%%%%%%%%%%%%%%%%%%%%%%%%%%%%%%%%%%%%%%%%%%%%%%%%%%%%%%%%%%%%%%%%%%%%%%%%%%%%%%%

A popular model for explaining X-ray plateaus observed in the afterglows of some gamma-ray bursts (GRBs) involves the formation of a millisecond magnetar that continuously pumps energy into the forward shock, thereby stalling the decay of the light curve \citep{zm01}. In many cases, however, the inferred polar-dipole strength needed to accommodate the observations is very large: \cite{rowl13} find that the plateau from GRB 070724A is best-fit with $B_{p} \approx 3 \times 10^{16}$~G, for example. Such values are difficult to reconcile with dynamo models, which have a hard time producing strong dipole components, favouring instead tangled multipoles\footnote{The dynamo simulations of \cite{rb22} find that the dipolar component contains $\sim 5\%$ of the total energy magnetic (though this may be underestimated due to numerical limitations on setting realistic Prandtl numbers). For a tangled field of average strength $\sim 10^{16}$~G we may thus expect $B_{p} \sim 5 \times 10^{14}$~G which, when accounting for GR spindown corrections, could be sufficient to accommodate the known Galactic magnetars.}. This event also displayed a sudden fall-off in the flux $\sim 90$~s after the main event, suggestive of a collapse time for a meta-stable, supramassive star [e.g., $(M/R)_{\rm max} \sim 0.3$ for many equations of state that pass multiwavelength constraints \citep{ofen24}].

The story changes somewhat if using instead the GR expression for the spindown luminosity \eqref{eq:sdlum}. For $M/R = 0.3$ we find $\kappa = 22.4$, meaning that the inferred field should be considerably lower. In fact, this is a lower limit since rapidly rotating stars have even further reductions: taking a $\sim 10\%$ change in the $tt$-component of the metric as representative (see footnote 1), we anticipate $N_R$ to decrease by a further $\sim 5\%$ so that $\kappa \approx 28$ for a $P \sim 2$~ms object \cite[see also figure 4 in][]{ruiz14}. Keeping the plateau duration fixed, \cite{rowl13} note that $B_{p}^2 \propto L^{-1}$ in their fitting procedure and thus -- all else being equal -- the inferred field for the putative star born in GRB 070724A may reduce to $B_{p} \lesssim 6 \times 10^{15}$~G or even lower with twists or rapid rotation. While still large, such a value is easier to accept \cite[for realistic energy conversion factors, the inferred $B_{p}$ decreases further; see][]{sk21}. 

By contrast, since the free energy of stable configurations is unchanged by compactness it may be harder to explain the `giant flares' with isotropic yields $\gtrsim 10^{50}$~erg observed up to $\sim 10^{5}$~s after the prompt emission in a number of plateau-exhibiting GRBs with a magnetar \cite[see][and references therein]{dab25}.

%%%%%%%%%%%%%%%%%%%%%%%%%%%%%%%%%%%%%%%%%%%%%%%%%%%%%%%%%%%%%%%%%%%%%%%%%%%%%%%%
\section{Conclusions}\label{sec:conclusions}
%%%%%%%%%%%%%%%%%%%%%%%%%%%%%%%%%%%%%%%%%%%%%%%%%%%%%%%%%%%%%%%%%%%%%%%%%%%%%%%%
%beyond that which we explore here. For example, \cite{beskin90} argued that pair creation may be possible over the larger portions of the polar cap in GR because of the stronger gravity and Lense-Thirring effect, as confirmed by recent simulations

In this paper, we have considered the problem of building GR and fully three-dimensional models of magnetar magnetospheres. While GR can influence several aspects of magnetar phenomena, such as pair cascades and particle drag dynamics \citep{beskin90,phil15}, our focus here is on the equilibrium of magnetic stresses. Using a prescription for the twist profile (expression~\ref{eq:alpha_surface}) based on the expectation that surface currents may develop in localised regions during the course of crustal evolution or otherwise, a sequence of models were considered of varying compactness and `hotspot' particulars (e.g. Fig.~\ref{fig:twisted_magnetospheres}). We find that the stable configurations (with respect to given boundary data) can house a free energy of at most $\sim 30\%$ regardless of the compactness (see Fig.~\ref{fig:energy_vs_alphamax_axisymmetric}), consistent with previous Newtonian studies \cite[e.g.][]{Akguen2018}. The free energy excess decreases in more realistic 3D cases with localised twists to at most $\approx 5\%$. Regardless of the twist particulars we also find that that $E_{e}$ is hardly changed as a function of compactness (see Fig.~\ref{fig:energy_vs_alphamax_3d}). We argue in Sec.~\ref{sec:discussion} that because the spindown luminosity scales strongly with GR effects \citep{Rezzolla2004}, the energy budget available to fuel magnetar bursts may be considerably lower than previously estimated. For some events -- notably the 2009 burst from SGR 0418+5729 \citep{van10} -- the dipolar excess may in fact be insufficient even for a (likely unrealistic) maximally-twisted configuration (see Fig.~\ref{fig:freevsburst}). This provides further evidence for multipolar components in magnetar fields. 

All solutions presented here were computed using a novel PINN solver (Sec.~\ref{sec:pinns}) and, to our knowledge, this is the first time such sequences have been constructed. This work thus highlights the value that machine-learning techniques have to offer in solving differential equations as, with relatively minimal expense, an ensemble function can be constructed that includes a wide parameter range \citep{Karniadakis2021}. This is especially useful in cases where such outputs may be fed into a different routine (see below) or where one wishes to build a database of outputs to systematically compare models with observations since the training need only be completed once. Traditional grid-based methods, for instance, require reevaluation for each set of boundary conditions which can be expensive if the parameter space is large, as for the problem considered here due to the wealth of twist ($\alpha_*^{\rm max}, \theta_1, \phi_1, \sigma$) and spacetime ($M/R$) particulars.

A number of directions would be natural to extend the results of this work. One obvious limitation is the exclusion of rotation. While Galactic magnetars spin slowly and metric corrections are small, an important element described throughout Sec.~\ref{sec:discussion} concerns the twist-enhanced spindown luminosity which we cannot directly estimate from our output. It would be interesting, for instance, to include such effects using the realistic current distributions from expression \eqref{eq:alpha_surface} to compare with the results of \cite{ng25}, who found enhancements of up to $\sim 16$ in the luminosity for extreme twists in an axisymmetric setup, but in GR 3D cases where the light cylinder lies at distances appropriate for stars with $P \sim 10$~s. PINNs could be particularly-well suited for such applications \citep{Stefanou2023a} or non-linear problems involving quantum-electrodynamic phenomena associated with hotspots \citep{cai22}. Other aspects which could be naturally incorporated via changes to the boundary conditions include multipoles, stellar spacetimes containing scalar or other hair, spheroidal stars deformed by hydromagnetic forces, or couplings with time-dependent internal processes \cite[see, e.g.,][]{Urban2023}. Such directions are left to future work.

%%%%%%%%%%%%%%%%%%%%%%%%%%%%%%%%%%%%%%%%%%%%%%%%%%%%%%%%%%%%%%%%%%%%%%%%%%%%%%%%
%%%%%%%%%%%%%%%%%%%%%%%%%%%%%%%%%%%%%%%%%%%%%%%%%%%%%%%%%%%%%%%%%%%%%%%%%%%%%%%%
\section*{Acknowledgements}
We thank Jos{\'e} Carlos Olvera for valuable guidance on the RNS code and an anonymous referee for their constructive feedback, particularly for their suggestion that resulted in Appendix C.
We acknowledge support from the Prometeo excellence programme grant
CIPROM/2022/13 funded by the Generalitat Valenciana. We also acknowledge 
support through the grant PID2021-127495NB-I00 funded by MCIN/AEI/10.13039/501100011033 and by the European Union, as well as the Astrophysics and High Energy Physics programme of the Generalitat Valenciana ASFAE/2022/026 funded by MCIN and the European Union NextGenerationEU (PRTR-C17.I1).

%%%%%%%%%%%%%%%%%%%%%%%%%%%%%%%%%%%%%%%%%%%%%%%%%%
\section*{Data Availability}

All data produced in this work will be shared on reasonable request to the corresponding author.

%%%%%%%%%%%%%%%%%%%% REFERENCES %%%%%%%%%%%%%%%%%%

\bibliographystyle{mnras}
\bibliography{references}

%%%%%%%%%%%%%%%%%%%%%%%%%%%%%%%%%%%%%%%%%%%%%%%%%%

%%%%%%%%%%%%%%%%% APPENDICES %%%%%%%%%%%%%%%%%%%%%

\appendix

\section{Details of the numerical implementation}\label{app:numerical_details}

We describe here the hyperparameters used to configure the PINN solver. The network is divided into four subnetworks, each dedicated to predicting one of the four target functions $B_r, B_\theta, B_\phi,$ and $ \alpha$.
For axisymmetric models, each subnetwork comprises two hidden layers with 20 neurons per layer, while for 3D models, each subnetwork comprises three hidden layers with 40 neurons each.  
The training set consists of $N=2\times10^{4}$ points for the axisymmetric cases and $N=8\times10^{4}$ points for the 3D cases, with the points randomly sampled within $\mathcal{D}$ which extends from the stellar surface to radial infinity in compactified coordinates (see below).

To ensure adequate coverage in the entire domain, the training points are resampled every 700 iterations.
The training process begins with the Adam optimiser, used for up to the first 1000 iterations to approach the loss minimum. After this initial phase, we switch to the \textit{self-scaled Broyden} method, which achieves significantly higher accuracy (often by several orders of magnitude) compared to commonly used optimization algorithms \citep[Adam, BFGS, L-BFGS; see][for details]{Urban2025}. We train the models for up to a maximum of 8000 total iterations
A typical training takes $\sim10-20$ minutes on a NVIDIA H100 GPU with 80GB RAM or $\sim 40 - 60$ minutes on a NVIDIA GeForce GTX 1660 Ti GPU with 6GB RAM for each model.
In the 3D case, this approach demonstrates faster performance (roughly by a factor of 2) than the method used in \cite{Stefanou2023} for Newtonian cases, although a direct comparison is challenging due to differences in the underlying hardware (GPU versus CPU).

\begin{figure}
    \centering
    \includegraphics[width=0.5\textwidth]{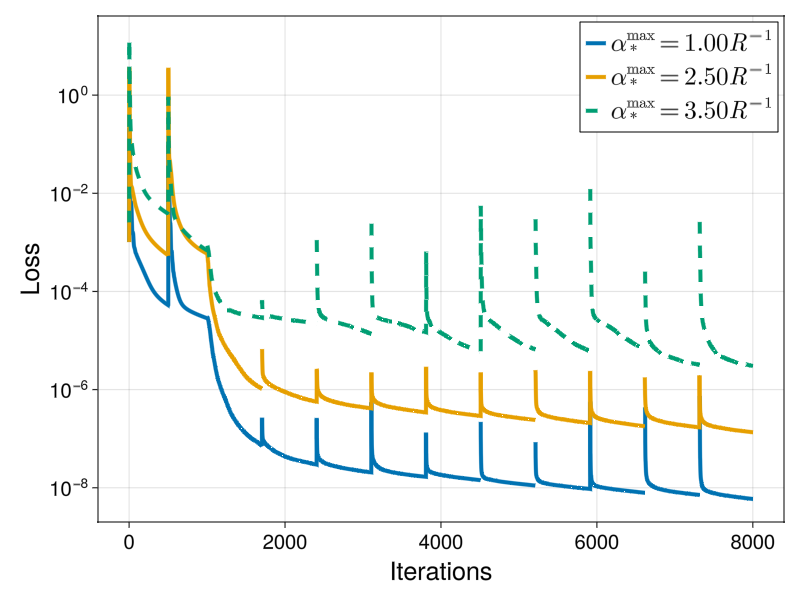}
    \caption{Convergence of the loss function with iterations for two relativistic 3D models with $M/R = 0.25$. The periodic peaks correspond to resamplings of the training points. The green dashed line shows an example of a model with poor convergence, which is discarded.}
    \label{fig:losses_vs_iterations}
\end{figure}

In all cases, we aim at a loss function that reaches values of at most $10^{-6}$, with the majority of models reaching $\mathcal{L} \lesssim 10^{-8}$ \cite[see][in particular their Appendix C, for a discussion about errors and accuracy in PINNs]{Urban2025}. 
The convergence plot of two relativistic 3D models is shown in Figure \ref{fig:losses_vs_iterations} as an illustrative example (orange and blue solid lines, same models as in the third column of Fig.~\ref{fig:twisted_magnetospheres}). We see that the highly twisted model ($\alpha_*^{\rm max} = 2.50 R^{-1}$), which is at the limit of acceptable, stable solutions (see Appendix \ref{sec:islands}), reaches a loss value $\sim 10^{-7}$ as opposed to the low twist model ($\alpha_*^{\rm max} = 1.00 R^{-1}$) which reaches $\sim 10^{-8}$. Similar convergence  plots are obtained for each of the models presented in this work. Most of the models that are discarded due to violating the FF condition (see Appendix \ref{sec:islands} below) present high peaks and the loss reduction halts to a value higher than $10^{-6}$, reflecting the difficulty of the solver to find the correct solution. Such a case is represented by the green dashed line in the figure.

\begin{figure}
    \centering
    \includegraphics[width=\linewidth]{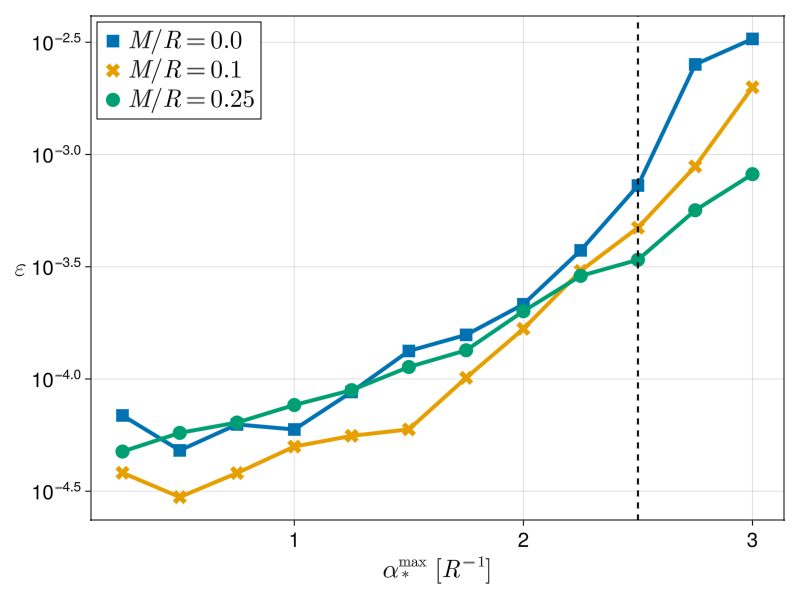}
    \caption{PDE error as a function of the maximum twist density at the surface for the 3D models constructed in Sec.~\ref{sec:3dresults}. The dashed line marks the value $\alpha_*^{\rm max} \simeq 2.5 R^{-1}$, beyond which the error reaches values $\gtrsim 10^{-3}$, indicating that convergence falls below our specified threshold.}
    \label{fig:error_vs_alphamax_3d}
\end{figure}

The global error $\varepsilon$ in the PDE can be estimated as $\varepsilon \simeq \sqrt{\mathcal{L}}$ \citep{Urban2025}, corresponding to an estimated precision of $\varepsilon \lesssim 10^{-4}$ as can be seen in Figure~\ref{fig:error_vs_alphamax_3d}, which is deemed sufficient. Models with $\varepsilon \gtrsim 10^{-3}$ (to the right of dashed line in the figure) are discarded as they violate the FF condition (see Appendix~\ref{sec:islands}). It is worth highlighting that the global errors are typically smaller in cases with non-zero compactness for the same $\alpha_*^{\rm max}$.
The actual error in the solution is expected to be lower than the PDE error \cite[again we direct the interested reader to][]{Urban2025}.

For convenience, we solve the FF equations in compactified Schwarzschild coordinates $(q, \theta, \phi)$, where $q = 1/r, \mu = \cos{\theta}$.
There are two reasons for this choice. First, the coordinates of the training points are of order unity and their intervals are compact, which is optimal for PINN training.
Second, imposing boundary conditions at infinity becomes straightforward as radial infinity is represented by a single point, $q = 0$, instead of by a box at some some large radius.
Additionally, we set $R = 1$ in order to simplify the expressions.
Written explicitly in these coordinates, our system of equations looks as follows: 
\begin{align}
    \frac{q \mu}{\sqrt{1 - \mu^2}} B_\phi - q \sqrt{1 - \mu^2} \parder{B_\phi}{\mu} - \frac{q}{\sqrt{1 - \mu^2}} \parder{B_\theta}{\phi} - \frac{\alpha B_r}{e^{\nu}} &= 0 \label{r_equation} \\
    \frac{q}{\sqrt{1 - \mu^2}} \parder{B_r}{\phi} - \frac{e^{-2\lambda} +1}{2 e^{-\lambda}} q B_\phi + e^{-\lambda} q^2 \parder{B_\phi}{q} - \frac{\alpha B_\theta}{e^{\nu}} &= 0 \label{theta_equation} \\
    \frac{e^{-2\lambda} +1}{2 e^{-\lambda}} q B_\theta - e^{-\lambda} q^2 \parder{B\theta}{q} + q \sqrt{1 - \mu^2} \parder{B_r}{\mu} - \frac{\alpha B_\phi}{e^{\nu}} &= 0 \label{phi_equation} \\
    -e^{-\lambda} q^2 B_r \parder{\alpha}{q} - q \sqrt{1 - \mu^2} B_\theta \parder{\alpha}{\theta} + \frac{q}{\sqrt{1 - \mu^2}} B_\phi \parder{\alpha}{\phi} &= 0 \label{alpha_equation} \\
     2 e^{-\lambda} q B_r - e^{-\lambda} q^2 \parder{B_r}{q} -  q \sqrt{1 - \mu^2} \parder{B_\theta}{\mu} \hspace{2.3cm} \notag \\
    + \frac{q \mu}{\sqrt{1 - \mu^2}} B_\theta + \frac{q}{\sqrt{1 - \mu^2}} \parder{B_\phi}{\phi} &= 0.  \label{divB_equation}
\end{align}

Boundary conditions are imposed through hard-enforcement.
If $\mathcal{N}_r, \mathcal{N}_\theta, \mathcal{N}_\phi, \mathcal{N}_\alpha$ are the PINN outputs, then the corresponding physical variables are defined through the following expressions:
\begin{align}
    B_r (q, \mu, \phi; \bm{\Theta}) &= q \left[ B_r^*(\mu, \phi) + (1-q) \mathcal{N}_r(q, \mu, \phi; \bm{\Theta}) \right] \label{eq:hard_enforcement_Br} \\
    B_\theta (q, \mu, \phi; \bm{\Theta}) &= q \mathcal{N}_\theta(q, \mu, \phi; \bm{\Theta}) \label{eq:hard_enforcement_Btheta} \\
    B_\phi (q, \mu, \phi; \bm{\Theta}) &= q \mathcal{N}_\phi(q, \mu, \phi; \bm{\Theta}) \label{eq:hard_enforcement_Bphi} \\
    \alpha (q, \mu, \phi; \bm{\Theta}) &= q \left[\alpha_*(\mu, \theta) + h_*(q, \mu, \phi)\mathcal{N}_\alpha(q, \mu, \phi; \bm{\Theta})\right]. \label{eq:hard_enforcement_alpha}
\end{align}
Here, $h_*$ is a function that is zero at points along the surface where the magnetic field is positive.
There are many options to achieve this behaviour.
One simple and suitable choice for accurate results is
\begin{equation}
    h_*(q, \mu, \phi) = q - 1 + \min(B_*(\mu, \phi), 0)^2.
\end{equation}
The functions $B_r^*$ and $\alpha_*$ define the boundary values of $B_r$ and $\alpha$, respectively. 
The factor $q$ in \eqref{eq:hard_enforcement_Br} - \eqref{eq:hard_enforcement_alpha} ensures that the relevant functions are zero at radial infinity ($q=0$). Note that, with these choices, all the desired boundary conditions are exactly satisfied independent of the output of the neural network. The PINN is trained so that its outputs $\mathcal{N}_r, \mathcal{N}_\theta, \mathcal{N}_\phi, \mathcal{N}_\alpha$ make equations \eqref{eq:hard_enforcement_Br}--\eqref{eq:hard_enforcement_alpha} a solution to the original system of equations \eqref{r_equation}--\eqref{divB_equation}.

%%%%%%%%%%%%%%%%%%%%%%%%%%%%%%%%%%%%%%%%%%%%%%%%%%%%%%%%%%%%%%%%%%%%%%%%%%%%%%%%%%%%%%%%%%%%%%%%%%%%
\section{Magnetic energies} \label{sec:magenergy}
%%%%%%%%%%%%%%%%%%%%%%%%%%%%%%%%%%%%%%%%%%%%%%%%%%%%%%%%%%%%%%%%%%%%%%%%%%%%%%%%%%%%%%%%%%%%%%%%%%%%

One of our aims in this paper is to determine the available magnetic energy stored in a twisted magnetosphere. In order to not distract from the flow of text, relevant details and definitions are provided in this Appendix, including an energy-specific method to check the validity of some numerical, FF output (Sec.~\ref{sec:islands}).

For all models, the magnetic energy is defined through
\begin{equation}\label{eq:energy}
    E^{(M)} = \int_V \frac{B^2}{8\pi}r^2 \sin \theta dr d \theta d \phi,
\end{equation}
which depends implicitly on the stellar mass. In the untwisted case, we find the analytic expression
\begin{equation} \label{eq:e0}
    \begin{aligned}
        E^{(M)}_{0} =& \frac{3 B_{*}^2 R^5 }{32 M^6} \Big[ 2 M (M + R) + R^2 \log\left(1-\frac{2M}{R}\right) \Big] \\
        & \times \Big[ 2 M  (M - R) + R (2 M - R) \log\left(1-\frac{2M}{R}\right) \Big],
    \end{aligned}
\end{equation}
where we have used the solutions \eqref{eq:br0} and \eqref{eq:bt0}. For example, one finds $E_{0}^{(0.25)} = 0.744 B_{*}^2 R^3$. In the Newtonian limit, $M \rightarrow 0$, expression \eqref{eq:e0} takes the expected value $E_0^{(0)} = B_{*}^2 R^3/3$; it increases as a function of the stellar compactness to leading-order as $E^{(M)}_{0} = B_{*}^2 R^3/3 \left[ 1 + 5 M / 2 R + \mathcal{O}(M^2) \right]$ \citep{Kojima2017}. We emphasize that these expressions apply for a given $B_*$: while valid for any (sub-Buchdahl) compactness, the inferred dipole moment decreases as a function of $M/R$ for given timing data (cf. equation \ref{eq:breduc} and Fig.~\ref{fig:ratios}).

\begin{figure}
    \centering
    \includegraphics[width=0.5\textwidth]{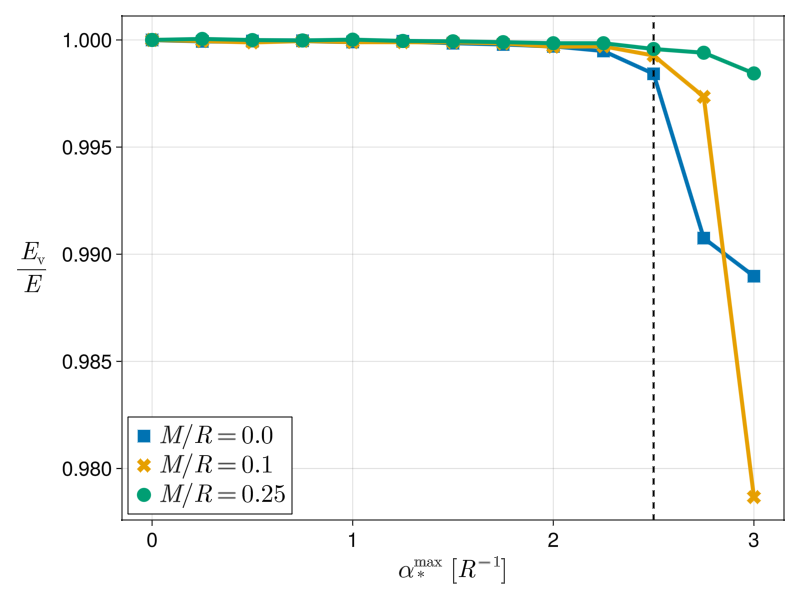}
    \caption{Ratio of the two energy definitions (equations \eqref{eq:energy} and \eqref{eq:lowintegral}) as a function of the maximum twist density at the surface for the 3D models of section \ref{sec:3dresults}. The dashed line marks the value $\alpha_*^{\rm max} \simeq 2.5 R^{-1}$, beyond which the ratio deviates from unity (although only in a 1-2 \%, note the scale), indicating that the FF condition is slightly violated.}
    \label{fig:virial_vs_alphamax_3d}
\end{figure}

%%%%%%%%%%%%%%%%%%%%%%%%%%%%%%%%%%%%%%%%%%%%%%%%%%%%%%%%%%%%%%%%%%%%%%%%%%%%%%%%%%%%%%%%%%%%%%%%%%%%
\subsection{Criteria to detect force-free breakdown} \label{sec:islands}
%%%%%%%%%%%%%%%%%%%%%%%%%%%%%%%%%%%%%%%%%%%%%%%%%%%%%%%%%%%%%%%%%%%%%%%%%%%%%%%%%%%%%%%%%%%%%%%%%%%%

A useful diagnostic for whether a given numerical output is strictly FF comes from the virial theorem \citep{c61}. It implies that a FF field within a (semi-infinite) domain $V$, decaying sufficiently fast with radius [$B \lesssim \mathcal{O}(r^{-3/2})$ at infinity], satisfies 
\begin{equation} \label{eq:lowintegral}
    \begin{aligned}
         E^{(M)}=& \frac{e^{2\nu(R)} R^3}{8\pi} 
        \int_{\partial V} \left( 
        B_{r}^2-B_{\theta}^{2}-B_{\phi}^{2}\right)\sin\theta d\theta d\phi \\
        & +\frac{1}{8\pi} \int_{V} 
        \left(1-e^{2\nu}\right)\left(B^2 + B_{r}^2\right) r^2\sin\theta
        dr d\theta d\phi.
    \end{aligned}
\end{equation}
Note that expression \eqref{eq:lowintegral} generalises the GR, axisymmetric result of \cite{yu11} and has not appeared in the literature before. In the limit $M \to 0$, the final volumetric piece vanishes identically and one recovers the \cite{low86} integral. In particular, a numerical mismatch between the left- and right-hand sides of \eqref{eq:lowintegral} indicates that the FF assumption has been violated somewhere in the domain \cite[provided $V$ satisfies some topological conditions; see][]{aly84}. 

One aspect of PINN solvers is that they can produce best-fit `solutions' even in cases where the equations cannot be strictly satisfied everywhere because of mathematical restrictions. 
PINNs are local solvers and are able to surpass these pathologies and produce a `correct' solution in the rest of the domain without crashing (but at the cost of increased loss function). 
This is an important advantage of this method, as it allows us to examine the domain of existence; a crash cannot, by itself, allow one to discern the non-existence of a solution.
Any deviation from unity from the ratio of the left- to right-hand sides of \eqref{eq:lowintegral} can be interpreted as a pathology in the magnetospheric configuration \cite[see also][]{klim92}. In this sense, it can be used -- together with the loss function \eqref{eq:loss} -- as a diagnostic for physically relevant solutions. 

Figure~\ref{fig:virial_vs_alphamax_3d} shows the ratio of the relevant expressions, which are of order unity until $\alpha_{*}^{\rm max} \approx 2.5 R^{-1}$. After this point some visible scatter appears meaning that, although the PINN solution is still optimal in some appropriate sense, it does not adequately satisfy the FF condition everywhere. It is for this reason we have considered models with $\alpha_{*}^{\rm max} \lesssim 2.5 R^{-1}$ in the main text (the exact maximum depends on twist particulars and compactness). Curiously, a significant deviation from unity starts where the models in \cite{Stefanou2023} stopped converging (for $M=0$) using an independent, Grad-Rubin method. For the axysymmetric models, the same logic was applied to select $\alpha_{*}^{\rm max} \lesssim 4 R^{-1}$.

\section{Stability and safety factor}

A critical factor in evaluating the stability of twisted magnetic fields is the so-called safety factor, which is defined as 
\begin{equation} \label{eq:safety}
    q_{\rm{sf}} = \frac{4 \pi}{\alpha \ell},
\end{equation}
for a field line (or a thin bundle) with constant twist density $\alpha$ and length $\ell$ \citep{Mahlmann2023, Rugg2024}. 
The inverse of the safety factor approximately indicates how many times magnetic field lines spiral around a flux tube. When the safety factor falls to about one or lower, the twisted magnetic structure becomes prone to kink instabilities \citep{ks54}, potentially triggering magnetar flares or outbursts
accompanied by rapid magnetic energy release. Thus, the safety factor can be used as a diagnostic tool to predict when a magnetospheric configuration becomes unstable.

\begin{figure}
    \centering
    \includegraphics[width=0.5\textwidth]{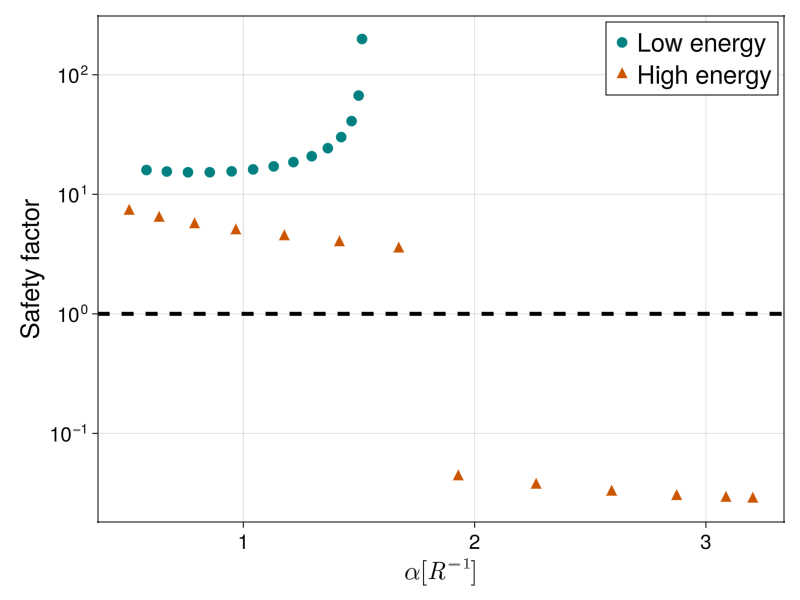}
    \caption{Field line safety factor $q_{\rm{sf}}$ (expression~\ref{eq:safety}) as a function of the twist density $\alpha$ for the degenerate axisymmetric solutions presented in Fig.~\ref{fig:alpha_contour_lines} ($M/R=0.25$, $s = 0.243$ and $m = 3$, resulting in $\alpha_*^{\rm{max}} = 1.86R^{-1}$). Only field lines with $\alpha > 0.5 R^{-1}$ and one of their footprints in the $\phi = \pi$ plane are shown for clarity. All field lines in the low-energy solution have $q_{\rm sf} \gg 1$, indicating stability. For the high-energy solution, a sudden jump in the safety factor to values $q_{\rm{sf}} \ll 1$ can be observed, which corresponds to the field lines that form the unstable magnetic island. Notice that for these lines $\alpha > \alpha_*^{\rm{max}}$.}
    \label{fig:safety_factor_axisymmetric}
\end{figure}

Axisymmetric models are known to be degenerate with the high-energy branch being unstable due to the presence of detached magnetic islands (see section \ref{sec:axisym} and Figure \ref{fig:energy_vs_alphamax_axisymmetric}). 
Indeed, as Figure \ref{fig:safety_factor_axisymmetric} shows, the safety factor of the field lines forming the magnetic island (after $\alpha_*^{\rm max} \approx 1.8$) in the high energy solution is $q_{\rm{sf}} \ll 1$, confirming that this region is highly susceptible to magnetic instabilities. Models in the low energy branch have a safety factor well above unity everywhere in the magnetosphere, as expected for the stable solutions.

\begin{figure}
    \centering
    \includegraphics[width=0.5\textwidth]{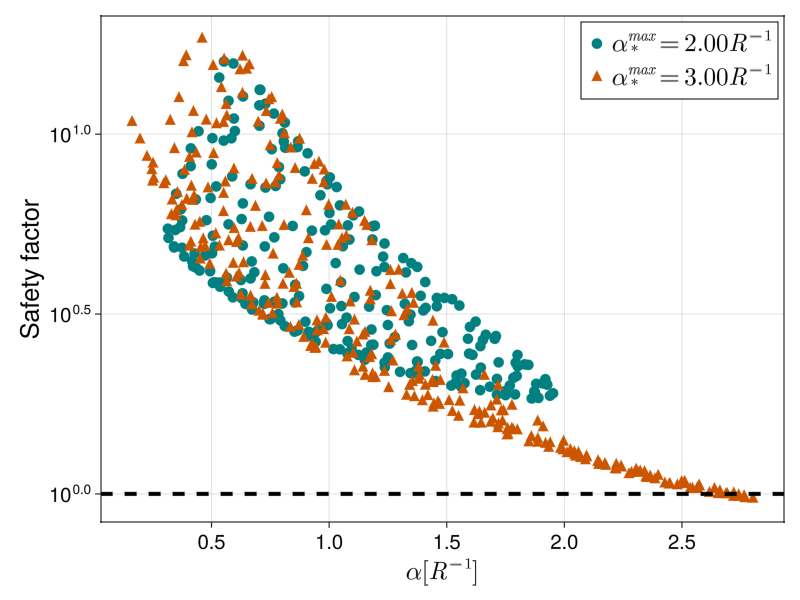}
    \caption{Safety factor $q_{\rm{sf}}$ as a function of the twist density $\alpha$ for a bundle of magnetic field lines emerging from the hotspot in two different relativistic 3D models: $\alpha_*^{\rm{max}} = 2 R^{-1}$
    (teal circles) and $\alpha_*^{\rm{max}} = 3 R^{-1}$ (orange triangles). Both models have compactness $M/R = 0.25$. Only field lines with $\alpha > 0.5 R^{-1}$ are shown for clarity.}
    \label{fig:safety_factor}
\end{figure}

For 3D models, a similar argument can be used. Figure \ref{fig:safety_factor} shows the safety factor of a bundle of twisted field lines for two relativistic 3D models. The less twisted model (teal circles in the figure) 
maintains a safety factor well above one for all its field lines.
Conversely, in the highly twisted model (orange triangles in the figure), a sub-bundle of field lines originating from the central region of the hotspot
has $q_{\rm{sf}} \lesssim 1$, suggesting the possibility for instabilities to develop. Remarkably, field lines that have $q_{\rm{sf}} \lesssim 1$ appear in models which have $\alpha_*^{\rm{max}} \gtrsim 2.5 R^{-1}$, the same value as the one that marks the breakdown of the FF condition according to Appendix \ref{sec:islands}. This reinforces the notion that beyond a critical twist, no stable FF solutions exist.

Relativistic corrections may work in favour of stability, albeit slightly. Increasing compactness has the effect of contracting flux bundles in the magnetosphere (see Figure \ref{fig:twisted_magnetospheres}), which results in higher safety factors due to the decreased length of the field lines. Indeed, for a fixed $\alpha_*^{\rm{max}}$, we have observed that $q_{\rm{sf}}$ is higher for compact stars, and may not reach the $q_{\rm{sf}} \sim 1$ threshold in cases where the Newtonian counterpart does. A similar effect is illustrated in Fig.~\ref{fig:virial_vs_alphamax_3d}, where the deviation from equality of the two energy definitions is slightly lower in the highly compact case. 
A detailed analysis of the connection between compactness and stability is outside the scope of this study and is planned for future research.

\bsp	% typesetting comment
\label{lastpage}

\end{document}